\documentclass[letterpaper, 10pt, conference]{ieeeconf}
\usepackage[T1]{fontenc}
\PassOptionsToPackage{hyphens}{url}
\usepackage{url}
\usepackage{algorithm}
\usepackage{algpseudocode}
\usepackage{graphicx}
\usepackage{array}
\usepackage{multirow}
\usepackage{booktabs}
\usepackage{pgfplots}
\usepackage{amsmath}
\usepackage{mdframed}
\usepackage{tikz}
\usetikzlibrary{arrows.meta, positioning, fit, backgrounds, calc}
\usepgfplotslibrary{fillbetween}
\pgfplotsset{compat=1.16}
\usepackage{minted}   % requires --shell-escape at compile time
\usepackage[
  pdftitle={Towards Persistent Case-Based Memory for Autonomous Data Science},
  pdfauthor={Felix Stocker},
  pdfsubject={Case-Based Reasoning, Autonomous Data Science, Small Language Models},
  pdfkeywords={case-based reasoning, autonomous data science, machine learning agents, small language models, Gemma 4},
  bookmarks=true,
  bookmarksopen=true,
  bookmarksnumbered=true,
  hidelinks
]{hyperref}

\IEEEoverridecommandlockouts
\newenvironment{IEEEkeywords}{%
  \par\vspace{0.5\baselineskip}\noindent
  \normalfont\small\textbf{\textit{Index Terms}}---\,\ignorespaces}{%
  \par\vspace{0.5\baselineskip}}

\title{\LARGE \bf
Towards Persistent Case-Based Memory for Autonomous Data Science:
A CBR-Augmented R\&D-Agent with a Locally Deployable Small Language Model
}

\author{%
  \authorblockN{Felix Stocker}
  \authorblockA{Technische Hochschule Ingolstadt (THI)\\
    Ingolstadt, Germany\\
    \texttt{fes4120@thi.de} $\,|\,$ \texttt{felix.stocker@sto-eng.de}}%
  \thanks{\textbf{Preprint.} This manuscript has not been submitted to or published in any journal or conference. May 2026.}%
}

\begin{document}

\maketitle
\thispagestyle{empty}
\pagestyle{empty}

%%%%%%%%%%%%%%%%%%%%%%%%%%%%%%%%%%%%%%%%%%%%%%%%%%%%%%%%%%%%%%%%%%%%%%%%%%%%%%%%
\begin{abstract}
Most top-performing autonomous data-science agents rely on frontier cloud models and lack persistent, cross-session memory.
This paper addresses two open gaps: (1)~the underexplored use of formally structured, quality-controlled Case-Based Reasoning (CBR) case bases coupling symbolic case records with executable code artefacts; and (2)~the untested viability of Small Language Models (SLMs) as locally deployable agent backbones.

This paper presents \emph{CBR-augmented R\&D-Agent}, integrating a persistent CBR layer into Microsoft's R\&D-Agent framework with a custom Google AI Studio backend that makes Gemma~4~31B~Dense operational as the agent backbone.
The CBR layer overrides three R\&D loop phases --- hypothesis generation, code generation, and case retention --- via a surgical subclass toggled by a single environment variable.
Cases are stored as structured records with executable code snapshots and quality metadata; a five-gate quality filter and a heuristic reuse-detection mechanism assess the plausibility of knowledge transfer by combining embedding similarity, code-fingerprint overlap, and injection provenance.
The Gemma~4 backend addresses the model's lack of native structured-output support through prompt-based schema enforcement, a state-machine JSON repair pipeline, and adaptive hang detection.

Evaluated on two Kaggle competitions (NOMAD~2018, Spaceship Titanic) with four seeds over eight improvement loops each, CBR achieves directionally higher accuracy than the CBR-disabled baseline (identical R\&D-Agent setup with CBR retrieval switched off) on Spaceship Titanic (0.8147 vs.\ 0.8098, $d=-1.41$) with substantially lower variance; on NOMAD the baseline's broader exploration yields a larger absolute SOTA gain.
Heuristic reuse detection across 108 retrieval events shows high semantic relevance (mean embedding similarity 0.882) alongside variable structural proximity (mean code-fingerprint similarity 0.305), consistent with conceptual guidance rather than verbatim code copying; establishing this directionally requires controlled ablation studies, which are deferred to future work.
The evaluation is a deliberate first step; cross-session transfer over the full MLE-Bench suite is the primary follow-on objective.
\end{abstract}


%%%%%%%%%%%%%%%%%%%%%%%%%%%%%%%%%%%%%%%%%%%%%%%%%%%%%%%%%%%%%%%%%%%%%%%%%%%%%%%%
\section{Introduction}

Automated machine learning and autonomous data-science agents have advanced rapidly since the introduction of LLM-driven pipelines.
Systems such as AIDE~\cite{aide}, MARS~\cite{mars}, Agent~K~\cite{agentk}, and the R\&D-Agent~\cite{rdagent} framework now compete on MLE-Bench~\cite{mlebench} at or beyond the level of human Kaggle competitors.
Yet two structural limitations persist across most published systems.

\emph{First}, memory typically remains transient: the dominant paradigm accumulates knowledge within a single run or competition and discards it when the session ends.
Among the systems that do address cross-session retrieval, the closest prior work is DS-Agent~\cite{dsagent}, which introduces an explicit CBR cycle on a fixed 30-task set, and Agent~K~\cite{agentk}, which employs a scaffold-derived chain-of-thought memory grounded in Kolb's experiential learning cycle while operating on a 72\,B-parameter backbone.
The specific design point of a \emph{formally structured, quality-controlled case base whose records couple symbolic case data with executable code artefacts and are designed to persist across competition boundaries} is, to my knowledge, comparatively underexplored.

\emph{Second}, the viability of SLMs as locally deployable autonomous agent backbones has not been demonstrated on competitive benchmarks.
The MLE-Bench baseline~\cite{mlebench} used o1-preview; subsequent leaderboard entries continue to rely on frontier cloud models.
The Gemma~4~31B~Dense model~\cite{gemma4card}, released in April 2026, reports dramatically improved agentic benchmarks over its predecessor --- but no published end-to-end MLE-agent evaluation has appeared.
This design point is particularly relevant for deployment contexts with limited GPU budgets, data-privacy requirements, or the need for transparent and maintainable system components --- conditions characteristic of small and medium-sized engineering teams.

\textbf{Scope and intent of this preprint.}
This paper is submitted as a preprint and should be read accordingly.
Its primary intent is threefold: to document the state of the art with particular attention to memory and local deployability; to establish that the proposed architecture is operational and produces plausible results; and to stake a design claim that structured, persistent case memory with a locally deployable open-weight model is a practically motivated direction distinct from the frontier-cloud-first paradigm.
The evaluation is deliberately limited: it shows the system runs correctly and yields interpretable results, not that it surpasses state-of-the-art systems or that all design claims are rigorously verified.
Limitations are detailed in Section~\ref{sec:discussion}; stronger evidence and cross-session evaluation are the subject of the follow-on roadmap defined there.

This paper makes three contributions:
\begin{enumerate}
\item A CBR layer for R\&D-Agent that provides persistent, quality-annotated case memory whose KB structure is designed to operate across competition domains, realised via a surgical three-method subclass.
\item A custom backend that makes Gemma~4~31B viable as an autonomous data-science agent backbone through two targeted engineering adaptations.
\item A first-step empirical evaluation on two Kaggle competitions with heuristic reuse detection, deliberately scoped to within-competition cases; cross-session evaluation requires a broader case base and is the primary follow-on target.
\end{enumerate}

The remainder of the paper is organised as follows. Section~\ref{sec:sota} reviews related work. Section~\ref{sec:method} describes the architecture. Section~\ref{sec:eval} presents the experimental setup and results. Section~\ref{sec:discussion} discusses findings and limitations. Section~\ref{sec:conclusion} concludes.

%%%%%%%%%%%%%%%%%%%%%%%%%%%%%%%%%%%%%%%%%%%%%%%%%%%%%%%%%%%%%%%%%%%%%%%%%%%%%%%%
\section{State of the Art}
\label{sec:sota}

\subsection{Case-Based Reasoning: Foundations and Modern Extensions}

Case-Based Reasoning (CBR) is conventionally organised around a four-step cycle of retrieval, reuse, revision, and retention, as formalised by Aamodt and Plaza~\cite{aamoplaza}.
The maintenance side of CBR --- the question of how cases may be removed or compressed without eroding the system's problem-solving capability --- was taken up by Smyth and Keane~\cite{smythkeane}, who introduced two deletion strategies that account for the competence of the case base rather than treating it as a pure performance bottleneck.
Together, these contributions establish the procedural and the housekeeping foundations of classical CBR.

The integration of CBR with transformer-based language models has attracted substantial recent attention.
Lewis et al.~\cite{rag} introduced Retrieval-Augmented Generation (RAG) as a general fine-tuning recipe in which a parametric language model is paired with a non-parametric retrieval index.
Wiratunga et al.~\cite{cbrrag} explicitly position the retrieval stage of RAG as a realisation of the CBR retrieval step: their CBR-RAG architecture for legal question-answering treats indexing vocabulary and similarity knowledge as first-class CBR containers and reports that this structured framing yields more contextually appropriate retrievals than a generic vector search.
Hatalis et al.~\cite{hatalis} review the CBR--LLM intersection in detail and observe that LLM agents face recurring difficulties with hallucination, contextual memory across interactions, and explanation transparency --- three areas where CBR mechanisms offer an established response.

Two recent community efforts crystallise the research agenda at this intersection.
The research manifesto by Bach, Floyd, Leake and colleagues~\cite{manifesto} proposes episodic memory as a means of enhancing LLM retention and discusses, in the opposite direction, the use of LLMs to improve CBR outputs.
Floyd and Leake~\cite{floydleake} propose a four-level taxonomy of agent memory --- short-term working memory, session memory, long-term memory, and human-like memory\allowbreak{} --- and state that current-generation LLMs operate primarily at the first two levels, with at most isolated elements of the third; they argue that a CBR case base, which can be added to and retrieved from indefinitely, is a natural realisation of long-term memory.
Bergmann et al.'s EXAR architecture~\cite{exar} develops this perspective into a unified design for experience-grounded agentic reasoning, positioning CBR as a memory and control structure within neuro-symbolic agentic systems and pointing to existing toolkits --- CBRkit and ProCAKE --- as concrete substrates.

A growing body of work demonstrates that the CBR cycle generalises productively to executable artefacts and structured knowledge engineering.
Dannenhauer et al.~\cite{cbrfewshot} use case retrieval for dynamic few-shot prompting in code generation and report higher accuracy than static three-shot and zero-shot prompting.
Brand et al.~\cite{brand} employ LLMs to generate both the vocabulary and the case base for a CBR system, addressing the long-recognised knowledge-acquisition bottleneck.
Minor and Kaucher~\cite{minor} consider a case base as a memory inside a RAG system for generating explanations of business process models.
Domain-specific instantiations span medicine~\cite{queipollano}, ecological restoration~\cite{ghazouani}, loan approval with counterfactual explanations~\cite{salimi}, note-taking workflows~\cite{craig}, and the evaluation of LLMs' decision-making capabilities~\cite{weber}.
Particularly germane to the present work, Zhou et al.'s Memento~\cite{memento} formalises agentic reasoning as a Memory-augmented Markov Decision Process equipped with a neural case-selection policy trained via soft Q-learning, stores past experiences as raw \emph{(state, action, reward)} tuples in an episodic case bank, and is explicit that the underlying LLM is not fine-tuned. Memento is instantiated for general-purpose deep-research tasks (GAIA, DeepResearcher) rather than autonomous data science, and its case bank retains all incoming cases without a quality filter --- the paper explicitly acknowledges the classic \emph{swamping problem} (where retrieval cost outweighs utility as the case bank grows unchecked) as a known challenge for this class of systems. The distinguishing design choices of the present work --- structured symbolic case records coupled with executable code artefacts, a five-gate quality filter, embedding-based deduplication, and a heuristic reuse-detection mechanism --- directly address this swamping challenge and are not present in Memento's design.

\begin{mdframed}

Together these contributions establish CBR as an actively developing framework for giving LLM agents persistent, explainable, structured experience.

Despite this momentum, two specific applications of CBR remain underexplored.
First, the use of CBR to couple runnable code artefacts with structured, quantitative quality signals (validation scores, improvement deltas, error traces, embedding-based deduplication, and heuristic reuse detection) for cross-session, cross-domain reuse in autonomous data science agents has, to my knowledge, not been demonstrated: DS-Agent~\cite{dsagent} (discussed in detail in Section~\ref{sec:sota_agents}) stores executable Python scripts in its deployment-stage case bank, but retains cases by a binary performance-improvement criterion only and does not annotate them with structured quality metadata.
Second, no prior work combines a formally structured CBR case base with a locally deployable small language model in this setting, leaving the question of whether the gap to frontier-model agents can be bridged through structured memory rather than parameter scaling entirely open.

\end{mdframed}

\subsection{Autonomous Data Science Agents}
\label{sec:sota_agents}

The vision of automated machine learning predates the LLM era.
Auto-WEKA~\cite{autoweka} performs the simultaneous selection of a learning algorithm and the tuning of its hyperparameters using Bayesian optimisation.
Auto-sklearn~\cite{autosklearn} builds on that pipeline and adds meta-learning to warmstart the search and automatic ensembling at the end.
TPOT~\cite{tpot} takes a different route and uses genetic programming to evolve tree-structured pipelines.
These systems excel within pre-specified search spaces but cannot reason about novel problem formulations or generate end-to-end code.

LLM-driven agents lift this restriction through natural-language planning and code generation.
Early systems such as Data Interpreter~\cite{datainterpreter}, MetaGPT~\cite{metagpt}, MLAgentBench~\cite{mlagentbench}, and AutoKaggle~\cite{autokaggle} established the basic pattern: decompose a complex task into subtasks, assign these to one or more LLM-driven roles, execute code in a runtime environment, and iteratively refine.
R\&D-Agent~\cite{rdagent} organises this loop explicitly into a research phase (planning, exploration-path structure, memory context, and reasoning pipeline) and a development phase (coding workflow and evaluation strategy), with a memory component that accumulates knowledge across iterations.
MLE-STAR~\cite{mlestar} addresses the observation that an agent relying solely on its LLM's internal knowledge tends to choose outdated model architectures: it invokes web search at the solution-initialisation stage to retrieve potentially state-of-the-art components and then performs ablation-guided refinement on individual code blocks.
AutoKaggle~\cite{autokaggle} decomposes a Kaggle competition into six phases --- background understanding, preliminary exploratory data analysis, data cleaning, in-depth exploratory data analysis, feature engineering, and a combined model-building, validation, and prediction stage --- executed by five specialised agents (Reader, Planner, Developer, Reviewer, Summarizer) and supported by a validated tool library.

DS-Agent~\cite{dsagent} is the system most directly aligned with the present work.
Guo et al.\ embed the four-step CBR cycle inside an LLM agent that operates in two stages: a development stage that retrieves textual expert insights from a Kaggle-derived human insight case bank to formulate experiment plans, and a deployment stage that retrieves and directly adapts executable Python scripts from an agent-generated case bank for code generation.
Across 30 selected data-science tasks (12 development, 18 deployment), DS-Agent with GPT-4 reaches a 100\% success rate across the 12 development tasks; in its deployment stage, transferring cases to weaker LLM backbones yields a 36\% average improvement in one-pass rate compared with non-CBR baselines~\cite{dsagent}.
This is, to the best of my knowledge, the first explicit CBR cycle in an LLM-driven data-science agent and provides the conceptual blueprint for the case-base design.
However, DS-Agent operates on a fixed within-distribution task set of 30 tasks; while its deployment stage does store and reuse executable Python scripts, both stages lack structured quality metadata (validation scores, improvement deltas, embedding-based deduplication), heuristic reuse detection, and the multi-gate retention filter that distinguish the present work, and the system is not designed for cross-session, cross-domain reuse.

The MLE-Bench benchmark~\cite{mlebench}, introduced by Chan et al.\ and published at ICLR 2025, has become a standard evaluation framework for end-to-end Machine Learning Engineering Agents (MLE) agents.
It curates 75 ML-engineering-related Kaggle competitions and grades agents against the real-world human leaderboards under an experimental setup of 24 hours per competition attempt.
The original evaluation reports that the best-performing configuration --- OpenAI's o1-preview combined with AIDE scaffolding --- reaches at least a Kaggle bronze medal in 16.9\% of the 75 competitions~\cite{mlebench}.
The live MLE-Bench leaderboard has continued to evolve and is now dominated by frontier cloud models; because that leaderboard changes on a near-weekly basis and the present work targets a different design point --- locally deployable models with persistent, transparent memory --- only those systems deemed most relevant and supported by a citable publication are discussed in Section~\ref{sec:sota_memory}, and the live leaderboard is not reproduced in full.
The live MLE-Bench leaderboard, maintained at \url{https://www.mlebench.com/}, has evolved rapidly since the initial publication.
Notably, OpenAI has temporarily paused the acceptance of new leaderboard submissions in order to improve the systematicity of the evaluation protocol, to make results more comparable across submissions, and to reduce the risk of evaluation artefacts introduced by differences in experimental setup.
This pause underscores a broader challenge for the field: a growing number of entries on the leaderboard --- including several near the top --- are backed by closed-source systems for which no detailed technical publication is available at the time of writing.
In the absence of peer-reviewed documentation, it is not possible to verify how these systems handle cross-session memory, what retrieval mechanisms (if any) they employ, or to what extent their reported gains stem from architectural innovations rather than from undisclosed engineering choices or proprietary data.
For the purposes of this work, the comparative analysis is
therefore restricted to the most relevant systems with a citable publication and documented implementation.

A second point of differentiation concerns the intended deployment context.
The majority of top-performing MLE-Bench systems are designed to operate as high-throughput research engines powered by frontier cloud APIs, optimised for peak benchmark performance under conditions of essentially unlimited computational budget and centralised infrastructure.

\begin{mdframed}

The system presented in this paper pursues a deliberately different goal: a transparent, locally deployable agent that is accessible to small and medium-sized engineering teams without requiring frontier cloud subscriptions, specialised infrastructure, or deep ML-engineering expertise to operate and extend.
This system is not intended to be a universal platform on which large organisations can build integrable agent pipelines at scale; rather, it is designed to be comprehensible, auditable, and maintainable by the small, domain-expert teams that are most likely to benefit from autonomous data-science support in practice.
The architectural choices reported below --- a locally runnable open-weight model, a transparent on-disk case store, and a surgically minimal CBR integration --- are all direct consequences of this deployment philosophy.

\end{mdframed}

\subsection{Memory and Experiential Learning in MLE Agents}
\label{sec:sota_memory}

A second axis of differentiation among contemporary MLE agents concerns the structure and scope of their memory mechanisms.
Most systems fall into one of three categories: in-context trajectory memory, structured intra-run lesson pools, or cross-session knowledge bases.

\textbf{In-context trajectory memory.}
AIDE~\cite{aide} organises all historical solutions of a single run as a tree whose nodes correspond to Python scripts and whose edges represent improvement attempts; this organisation guides iterative refinement within a run but no mechanism for cross-session persistence is part of the published design.
OpenHands~\cite{openhands} provides a sandboxed runtime, a multi-agent delegation mechanism, and a shared workspace for collaboration; like AIDE, the memory it exposes is centred on intra-run state rather than persistent recall.

\textbf{Structured intra-run lesson pools.}
MARS~\cite{mars} introduces a \emph{Comparative Reflective Memory} that handles credit assignment by comparing the current solution with the best-known one and distilling the difference into explicit \emph{Debugging Lessons} and \emph{Solution Improvement Lessons}.
The agent applies these lessons at a measured rate of 65.8~$\pm$~1.1\%; 63\% of the applied lessons originate from a different branch of the search tree than the one in which they are reused, which the authors interpret as the system's ``Aha!'' moments --- long-horizon performance jumps triggered by the transfer of a single high-signal lesson~\cite{mars}.
R\&D-Agent~\cite{rdagent} maintains an Memory Context component that accumulates knowledge across the iterations of a single project --- where \emph{single project} denotes one autonomous run on one competition, lasting until the configurable loop budget is exhausted.
Within this horizon, the memory context stores historical hypotheses, their associated code, and evaluation outcomes and makes them available to the reasoning pipeline for subsequent iterations.
Knowledge transfer between loops within a run is mediated by CoSTEER (\emph{Collaborative Knowledge-STudying-Enhanced Evolution by Retrieval})~\cite{costeer}, the sub-agent responsible for code synthesis in R\&D-Agent.
CoSTEER builds its own fine-grained \emph{practical knowledge base} that archives the full implementation trace of every successfully completed coding task --- not only the final solution but also the iterative trial-and-feedback history --- and retrieves from this trace at the level of \emph{error similarity} rather than task similarity: when a new coding error is encountered, CoSTEER queries the trace store for prior errors that match the current failure mode and injects the corresponding fix as in-context guidance.
Structurally, this intra-run knowledge store differs from the persistent CBR case base introduced in the present work in two important ways.
First, it is scoped to a single run: while R\&D-Agent's framework allows its internal stores to be serialised to disk, no mechanism is provided for indexing, querying, or quality-filtering these records across competition boundaries, meaning accumulated knowledge is effectively inaccessible to subsequent runs on different tasks.
Second, the granularity of stored knowledge differs: CoSTEER's trace records capture implementation-level error--fix pairs, whereas the CBR case base introduced in the present work stores solution-level records that couple a structured problem signature, a solution summary, a full code snapshot, and quantitative quality metadata, making individual cases retrievable and interpretable at the level of complete solution strategies rather than individual debugging steps.

\textbf{Cross-session knowledge bases.}
Agent~K~\cite{agentk} grounds its design in Kolb's experiential learning theory and Vygotsky's zone of proximal development.
Its architecture distinguishes an extrinsic loop (environment interaction such as code execution and feedback) from an intrinsic loop (reflection and abstraction over the agent's internal state).
As a comparison baseline, the paper evaluates a RAG-augmented ReAct agent that queries a Kaggle-based RAG database constructed from 24 competitions that started in or after 2021 and were not part of the benchmark set; Agent~K itself relies on scaffold-derived chain-of-thought abstractions rather than external retrieval.
Across 81 tasks, Agent~K --- using a Qwen2.5-72B backbone --- reaches an Elo-MMR score of 1694, placing it beyond the median performance of the Kaggle Masters in the study.
The Kaggle Masters represent an elite group of fewer than 2\% of the platform's user base of over 200\,000.
The paper reports medal-equivalent performance corresponding to 9 gold, 8 silver, and 12 bronze medals overall, of which 4 gold and 4 silver were obtained on prize-awarding\allowbreak{} (featured and research) competitions; the remaining counts apply to competitions that did not officially award medals, scored using a standard benchmarking convention~\cite{agentk}.
ML-Master~2.0~\cite{mlmaster} introduces a Hierarchical Cognitive Caching (HCC) architecture that explicitly separates context into three tiers: intra-run evolving experience (L1), intra-run refined knowledge summaries (L2), and a persistent cross-task prior wisdom cache (L3). The L3 cache stores task-agnostic, transferable strategies distilled from previously solved tasks and persists across sessions; the system was pre-populated from 407 Kaggle competitions excluded from the benchmark. Among the surveyed systems, ML-Master~2.0 is unique in building its external store dynamically from its own task completions: after each completed task, the task-level promotion operator distils the agent's own experience into the L3 cache. By contrast, AutoMind's ~\cite{automind} knowledge base and Agent~K's RAG baseline are pre-built from externally curated human knowledge and remain static during agent operation. ML-Master~2.0's L3 therefore represents the closest analogue to the present work in terms of agent-generated, cross-session persistence --- though it stores distilled textual wisdom rather than structured symbolic case records coupled with executable code artefacts and quantitative quality metadata, and does not provide the multi-gate quality filter or the heuristic reuse-detection mechanism introduced here.
AutoMind~\cite{automind} takes a complementary route, combining a curated expert knowledge base, an agentic knowledgeable tree-search algorithm, and a self-adaptive coding strategy that tailors code generation to task complexity. AutoMind's knowledge base is pre-built from 3\,237 public forum posts spanning 455 Kaggle competitions and recent ML papers, and is static --- it is not grown dynamically from the agent's own task completions.
ML-Agent~\cite{mlagent} explores a different paradigm: it trains a 7\,B-parameter Qwen-2.5 LLM with online reinforcement learning on 9 ML tasks and reports that the resulting agent outperforms a 671\,B-parameter DeepSeek-R1 agent on 3 held-in and 10 held-out tasks --- showing that learned action policies can substitute for parametric scale.

\begin{mdframed}

What is, to my knowledge, not documented in the surveyed systems is a formally structured, quality-controlled, externally inspectable case base that (i)~persists across sessions, (ii)~generalises across domains rather than within a single competition family, (iii)~couples symbolic case records with executable artefacts, and (iv)~is designed for transparent extension and maintenance.
The RAG database used by Agent~K's ReAct+RAG baseline approximates (i) and (ii); its paper describes the database as built from Kaggle competition material, which sits at a different level of abstraction from structured cases with explicit quality metadata.
ML-Master~2.0's L3 Prior Wisdom cache also approximates (i) and (ii): it persists across tasks and was pre-populated from 407 diverse Kaggle competitions, but stores distilled textual wisdom rather than symbolic case records with executable code artefacts and quality metadata, and does not provide the structured retrieval and quality-filtering mechanisms targeted here.
DS-Agent's agent case bank most directly approximates (iii): it explicitly stores \emph{(task description, Python script)} pairs and reuses them for code generation in its deployment stage, but the case bank is scoped to a fixed within-distribution task set, carries no quality metadata beyond binary success, and provides no embedding-based deduplication or heuristic reuse detection.
MARS's lesson pool approaches the symbolic half of (iii) through its structured textual lesson records --- each containing an isolated causal change, a comparative impact analysis, and a generalised rule --- but does not store executable code artefacts; the lesson pool is constructed within a single competition's search tree.

\end{mdframed}

\subsection{Small Language Models: Capabilities and Challenges}

The computational and financial costs of frontier LLM inference have motivated sustained interest in smaller, locally deployable alternatives.
Touvron et al.'s LLaMA~\cite{llama} showed that careful training-data curation can yield comparatively small models with competitive performance: LLaMA-13B exceeds GPT-3 on most benchmarks while having roughly one tenth of its parameters.
Subsequent work on Mistral~7B~\cite{mistral} reports a 7\,B-parameter model that exceeds the Llama-2 13B baseline across all evaluated benchmarks and exceeds the larger Llama-1 34B specifically in reasoning, mathematics, and code generation.
DeepSeek-Coder~\cite{deepseekcoder} confirmed the viability of sub-30\,B models for programming tasks, releasing an open-source code-model series ranging from 1.3\,B to 33\,B parameters and reporting state-of-the-art performance among open-source code models on multiple benchmarks.

Despite these advances, SLMs have faced well-documented challenges in the autonomous agent setting.
Kambhampati et al.~\cite{kambhampati} argue, on the basis of consistent empirical evidence on planning benchmarks, that auto-regressive LLMs cannot plan or self-verify by themselves and are better positioned as components inside an LLM-modulo framework where an external model-based verifier handles plan validation.
A second class of failure mode for long-horizon agents is context degradation: as input contexts grow longer, language models increasingly struggle to robustly access and use relevant information. Performance declines particularly when task-relevant information is embedded in the middle of long contexts rather than at the beginning or end.~\cite{lostinthemiddle}.
These dynamics affect smaller models more strongly than frontier-scale models and contributed to the prevailing assessment, prior to the most recent generation of open-weight models, that SLMs were not suitable for end-to-end MLE-Bench-style evaluation.

\subsection{Gemma~4: A New Generation of Locally Deployable Models}
\label{sec:sota_gemma4}

Google DeepMind released the Gemma~4 family on 2~April~2026~\cite{gemma4blog}.
According to the official Gemma~4 model card~\cite{gemma4card}, the release comprises four open-weight variants intended to span deployments from mobile and edge devices to consumer GPUs and workstations: two small models, E2B (2.3\,B effective parameters) and E4B (4.5\,B effective parameters), each with a 128\,K context window; a Mixture-of-Experts variant labelled 26B~A4B with 25.2\,B total and 3.8\,B active parameters; and a 31\,B Dense variant with 30.7\,B parameters and a 256\,K context window.
The present work focuses on the 31\,B Dense variant, which is the highest-performing locally deployable option in the family.

The vendor-reported benchmark suite documented in the Gemma~4 model card~\cite{gemma4card} indicates substantial gains for the instruction-tuned 31\,B Dense variant over Gemma~3~27B in its non-thinking mode along every dimension relevant to agentic coding.
Reasoning and knowledge benchmarks move from 67.6\% to 85.2\% on MMLU~Pro, from 20.8\% to 89.2\% on AIME~2026 (no tools), and from 42.4\% to 84.3\% on GPQA~Diamond.
The coding side moves from 29.1\% to 80.0\% on LiveCodeBench~v6 and from a Codeforces ELO of 110 to one of 2150.
On the BigBench Extra Hard reasoning benchmark, the reported score rises from 19.3\% to 74.4\%.
On the $\tau^2$-bench agentic-tool-use benchmark --- the metric most directly diagnostic for autonomous agent behaviour --- the model card reports a Tau2 score averaged over three splits of 76.9\% for Gemma~4~31B compared with 16.2\% for Gemma~3~27B~\cite{gemma4card}.
As with any vendor-published evaluation, these figures should be read as the manufacturer's own numbers; independent third-party reproductions on MLE-Bench-style end-to-end agent workloads have not yet appeared in the literature.

\begin{mdframed}

Three capability changes documented in the official model card~\cite{gemma4card} and accompanying launch blog~\cite{gemma4blog} are particularly relevant for the present work.
First, Gemma~4 supports configurable thinking modes that let the model produce step-by-step reasoning before generating a final answer.
Second, structured function calling is offered as a native capability, so that an agent can emit JSON-structured tool calls without grammar-constrained generation.
Third, the 31\,B Dense variant carries a 256\,K context window, which directly addresses the cognitive-overload failure mode identified above and which, for a CBR-augmented pipeline, is large enough to accommodate a retrieved case together with its prior code, error traces, and quality annotations without truncation.
To the best of my knowledge, no prior published evaluation has examined Gemma~4 in the role of a backbone for an autonomous data-science agent, and no prior system has combined Gemma~4 with a persistent, quality-controlled CBR layer.

\end{mdframed}

\subsection{Research Gaps Addressed by This Work}

Two interrelated gaps motivate this study.
The design is first situated relative to the systems closest in goal --- DS-Agent, R\&D-Agent, Agent~K, MARS, AIDE, ML-Agent, and ML-Master~2.0 --- and then the gaps are stated formally.

\textbf{The closest prior systems.}
DS-Agent~\cite{dsagent} is, to my knowledge, the most directly aligned prior system in embedding an explicit CBR cycle in an LLM-driven data-science agent; it stores executable Python scripts in its agent case bank and reuses them directly in its deployment stage, but both stages lack structured quality metadata, embedding-based deduplication, heuristic reuse detection, and the multi-gate retention filter of the present work, and the system is scoped to a fixed within-distribution task set rather than designed for cross-domain, cross-session transfer.
R\&D-Agent~\cite{rdagent} organises agentic data-science work explicitly around hypothesis generation and a development phase, with an FC-Memory Context component --- where FC denotes \emph{Framework Concept}, the label used by Yang et al.\ for each of the six modular, independently replaceable components that define the framework's design space --- that, as documented in the framework, accumulates knowledge across the iterations of a single project.
Agent~K~\cite{agentk} grounds its architecture in Kolb's experiential learning cycle and Vygotsky's zone of proximal development, operating with a Qwen2.5-72B backbone; its scaffold-derived chain-of-thought abstractions guide open-ended solution generation rather than querying an external retrieval index at inference time. As a comparison baseline, the paper evaluates a RAG-augmented ReAct agent whose retrieval database was constructed from 24 Kaggle competitions that started in or after 2021 --- the spiritually closest analogue to the goal of the present work --- but this RAG component belongs to the baseline Agent~K outperforms, not to Agent~K itself; the database is described as being constructed from Kaggle competition material, which sits at a different level of abstraction from the layer of structured, externally inspectable case records with explicit quality metadata pursued here.
MARS~\cite{mars} maintains a Comparative Reflective Memory with a documented 63\% cross-branch lesson transfer, with the ``Aha!''\ phenomenon as a consequence; its lesson pool is constructed within a single competition's search tree, with a Review Agent that filters redundant lessons via LLM-based reasoning, though this redundancy check is not equivalent to the performance-based quality gates and heuristic reuse detection of the present work.
AIDE~\cite{aide} organises a single run's solutions in a tree structure and serves as the de facto scaffolding baseline in the field.
ML-Agent~\cite{mlagent} internalises cross-task knowledge through online reinforcement learning, showing that a 7\,B-parameter Qwen-2.5 agent can outperform a 671\,B-parameter DeepSeek-R1 agent on held-out tasks; the resulting knowledge, however, is parametric and not externally inspectable.

ML-Master~2.0~\cite{mlmaster} introduces a Hierarchical Cognitive Caching architecture with a persistent L3 Prior Wisdom cache that stores task-agnostic, transferable strategies across task boundaries; the system was pre-populated from 407 Kaggle competitions excluded from the benchmark, making it the only surveyed system besides the Agent~K RAG baseline with an externally populated, retrievable cross-task store. However, the L3 cache stores distilled textual wisdom obtained through LLM-based summarisation rather than structured symbolic case records coupled with executable code artefacts, and neither quality-gate filtering, embedding-based deduplication, nor heuristic reuse detection are part of its design.

\begin{mdframed}

{\tolerance=2000\par\noindent\textbf{Gap~1: A formally structured, quality-controlled CBR case base that couples symbolic case records with executable code artefacts and is designed for cross-session, cross-domain reuse remains comparatively underexplored in autonomous data-science agents.}\par}

\end{mdframed}

Among the systems above, cross-session and cross-task knowledge is approached through retrieval over Kaggle competition material (the ReAct+RAG baseline evaluated alongside Agent~K), a persistent distilled-wisdom cache (ML-Master~2.0), structured intra-run lesson pools (MARS, R\&D-Agent), parametric encoding of cross-task knowledge (ML-Agent), or CBR over a fixed task set (DS-Agent).
To the best of my knowledge, no surveyed system instantiates the specific combination targeted here: a persistent, externally inspectable, quality-annotated case base whose records couple symbolic case data with executable code artefacts and whose KB structure is explicitly designed to admit cases from different competition domains in the same store.

\begin{mdframed}

{\tolerance=2000\par\noindent\textbf{Gap~2: No systematic evidence on SLMs as the backbone of autonomous data-science agents.}\par}

\end{mdframed}

The original MLE-Bench evaluation~\cite{mlebench} established that scaffolding-based agents on frontier cloud models (o1-preview with AIDE) reached a bronze-medal level in 16.9\% of the 75 competitions, and subsequent leaderboard entries have continued to draw on frontier cloud backbones.
The pre-Gemma-4 viability assessment for smaller open-weight models was negative (Section~\ref{sec:sota_memory}), and the vendor-reported Gemma~4~31B benchmark gains documented in the preceding section suggest that this assessment is now outdated --- but the question has not been tested end-to-end on MLE-Bench in any published evaluation.

These two gaps are addressed jointly by the architecture presented in Section~\ref{sec:method}.

%%%%%%%%%%%%%%%%%%%%%%%%%%%%%%%%%%%%%%%%%%%%%%%%%%%%%%%%%%%%%%%%%%%%%%%%%%%%%%%%
\section{Methodology}
\label{sec:method}

\subsection{R\&D-Agent: Baseline Architecture}

R\&D-Agent~\cite{rdagent} is an open-source LLM-agent framework for autonomous data science.
It structures one improvement run as an iterative \emph{R\&D loop} of five sequential phases: (1)~hypothesis generation and experiment specification, (2)~code synthesis via the CoSTEER debug-iterate sub-loop, (3)~isolated code execution, (4)~result collection with an LLM-based quality judge, and (5)~an implicit retention step in which the framework decides whether to advance the SOTA pointer.
The loop repeats until a configurable budget is exhausted; a trace object accumulates hypotheses and outcomes within the session.
While R\&D-Agent serialises its internal stores to disk between runs, it provides no mechanisms to index, filter, or query these records across competition domains: re-using accumulated knowledge on a different competition requires manual intervention.

The five loop phases and the three CBR overrides are illustrated in Fig.~\ref{fig:rdagent_loop} (Appendix); Algorithm~\ref{alg:cbr_loop} provides the complete pseudocode.

\subsection{CBR Integration into R\&D-Agent}

\subsubsection{Design Rationale}

The CBR layer is realised as a subclass of R\&D-Agent's loop controller that overrides exactly three phases --- hypothesis generation, code synthesis, and retention --- leaving all remaining logic untouched.\footnote{Source code: \url{https://github.com/stofe94/cbr-rd-agent}}
This allows the layer to be toggled by a single environment variable without modifying upstream source files, which is essential for the A/B comparisons in Section~\ref{sec:eval}.

\subsubsection{Knowledge Base Structure}

Each entry in the knowledge base (KB) is a \emph{Case}, a structured record comprising three components: a \emph{problem signature} (task and data type, evaluation metric and direction, competition context), a \emph{solution signature} (hypothesis text, plan summary, inferred model family, key techniques, and a full code snapshot), and a \emph{case-metrics} record (primary metric value, baseline value, computed improvement).
Each case also carries provenance fields and per-case reuse counters updated by the heuristic reuse-detection mechanism described in Section~\ref{sec:causal_reuse}.

Vector retrieval is implemented on top of R\&D-Agent's existing vector-base component.
Each case is embedded from a compact structured string built only from problem and solution fields --- no free-form code or injected context enters the embedding input, so the retrieval index reflects problem--solution semantics rather than incidental textual overlap.
The KB persists as three on-disk artefacts: a case-metadata file, a code-snapshot directory (one Python file per case), and a serialised vector index.
A consistency check on initialisation re-embeds any case whose vector row is missing and prunes orphan vector rows, ensuring that a crash between two writes cannot produce a permanently divergent store.

\subsubsection{Retrieval and Context Injection}

CBR context is injected at two points in the R\&D loop.

\textbf{Hypothesis phase.}
Before hypothesis generation, the CBR layer performs a pre-retrieval against the current scenario description to warm the embedding cache.
After the parent generates a hypothesis, a second retrieval is performed using the full hypothesis text combined with scenario metadata as the query, yielding up to a configurable number of cases with embedding similarity above a configurable threshold.
Retrieved cases and any relevant failure patterns from the Failure Tracker are formatted as a structured Markdown block --- containing approach description, key techniques, and a truncated plan summary --- and written into the hypothesis's appendix field.
The appendix field was chosen over the main reason field so that the plan summary stored in the KB is built from the clean concise fields of the hypothesis rather than from CBR-injected text.

\textbf{Coding phase.}
The coding override iterates over the experiment's task descriptions and appends a formatted code-reference section to each, containing the full code snapshot of each retrieved case together with a metadata header (similarity score, task type, data type, model family, achieved metric).
Additionally, a CBR reference document is written to the experiment workspace so that the CoSTEER sub-loop can access it across its internal debug iterations.

Algorithm~\ref{alg:cbr_loop} gives the pseudocode for one complete loop iteration.
Phases~1 and~2 (hypothesis generation and code synthesis) are CBR overrides: a two-stage retrieval injects relevant cases and Failure-Tracker patterns as structured context into both the hypothesis and the coding prompt.
Phases~3 and~4 (code execution and LLM-based feedback) run unchanged.
In Phase~5 (retention), the Quality Gate decides whether to add the experiment to the KB or forward it to the Failure Tracker as negative guidance; \textit{UpdateReuseCounters} then records which injected cases contributed to a structurally similar new solution.

\begin{algorithm}[t]
\scriptsize
\caption{CBR-augmented R\&D-Agent Loop}
\label{alg:cbr_loop}
\begin{algorithmic}[1]
\State \textbf{In:} $S$ (scenario), $T$ (trace), $\mathit{KB}$, $\mathit{FT}$, $L$ (budget)
\State \textbf{Out:} Updated $\mathit{KB}$, best metric $m^*$
\For{$\ell = 1$ \textbf{to} $L$}
  \State \textit{// Phase 1 — hypothesis generation}
  \State $\mathit{cases} \leftarrow \mathit{KB}.\texttt{retrieve}(S)$ \hfill{\scriptsize\textit{pre-retrieval}}
  \State $h \leftarrow \texttt{GenerateHypothesis}(T,\,S)$ \hfill{\scriptsize\textit{R\&D-Agent}}
  \State $\mathit{cases} \leftarrow \mathit{KB}.\texttt{retrieve}(h,\,S)$
  \State $h \leftarrow \texttt{EnrichWithContext}(h,\,\mathit{cases},\,\mathit{FT})$
  \State \textit{// Phase 2 — code generation}
  \State $\mathit{exp} \leftarrow \texttt{GenerateCode}(h,\,\mathit{cases})$ \hfill{\scriptsize\textit{CoSTEER}}
  \State \textit{// Phases 3\,\&\,4 — execution \& feedback}
  \State $r \leftarrow \texttt{Execute}(\mathit{exp})$;\enspace
         $\mathit{fb} \leftarrow \texttt{EvaluateFeedback}(\mathit{exp},\,r)$
  \State \textit{// Phase 5 — retention}
  \State $c \leftarrow \texttt{BuildCase}(h,\,\mathit{exp},\,r,\,S,\,m^*)$
  \If{$\texttt{QualityGate}(c,\,\mathit{fb},\,\mathit{KB})$}
    \State $\mathit{KB}.\texttt{addCase}(c)$;\enspace $m^* \leftarrow c.\mathit{metric}$
    \State $\texttt{UpdateReuseCounters}(c,\,\mathit{cases},\,\mathit{KB})$
  \Else
    \State $\mathit{FT}.\texttt{log}(c)$
  \EndIf
\EndFor
\end{algorithmic}
\end{algorithm}

\subsubsection{Five-Gate Quality Filter}

Five sequential quality-gate checks determine whether an experiment is retained in the KB.
\textbf{G1}~checks execution success by scanning the execution log for runtime errors.
\textbf{G2}~verifies that a numeric primary metric value is extractable from the log via regex-based metric-alias resolution.
\textbf{G3}~checks whether the result improves over the current SOTA baseline in a direction-aware manner; it is skipped on the first run when no baseline exists yet.
\textbf{G4}~ensures novelty by checking that the nearest neighbour in the KB falls below an embedding-similarity deduplication threshold, unless the candidate is strictly better than its nearest neighbour.
\textbf{G5}~makes an LLM call to extract transferable model type and key techniques from the experiment, enriching the case metadata unconditionally without ever rejecting a case.

Cases failing G1 or G2 are treated as hard failures and forwarded to the Failure Tracker, which stores them in a separate embedding index so that subsequent retrieve--inject cycles can warn the agent about known bad approaches.
Cases failing G3 or G4 are discarded silently.

\subsubsection{Heuristic Reuse Detection}
\label{sec:causal_reuse}

After every successful retention, a heuristic reuse-detection check assesses plausibility of knowledge transfer by testing two conditions jointly: (i)~embedding similarity between the new and source case exceeds the retrieval threshold; and (ii)~code-fingerprint similarity (normalised Jaccard trigram overlap) exceeds a structural threshold.
A positive reuse signal is credited only for cases actually injected into the coding prompt, excluding coincidental similarity with non-injected cases.
The per-case heuristic-reuse rate computed from this mechanism is stored in the KB and surfaced in the retrieval prompt header, giving the model a proxy quality signal about the track record of individual cases.
Specific threshold values are listed in the experimental configuration (Section~\ref{sec:exp_config}).

\subsection{Google AI Studio Backend for Gemma~4}

A custom Google AI Studio backend implements R\&D-Agent's LLM interface contract against Google's official SDK. Compared with frontier models such as Gemini~2.5~Pro, integrating Gemma~4~31B required two engineering adaptations targeting specific failure modes observed during development.

\textbf{Structured output without native schema support.}
R\&D-Agent parses LLM responses against Pydantic schemas throughout its pipeline --- for hypothesis objects, experiment plans, coding task descriptions, and quality judgements.
Gemma~4's native function calling covers standard tool invocations but not the schema-constrained structured output required throughout R\&D-Agent's pipeline; the backend compensates through prompt-based schema enforcement and a multi-stage repair pipeline.
Before every structured request, a Gemma-4-specific template is injected into the system message, making the target JSON schema explicit and stating concrete formatting rules.
The schema is compressed by stripping non-structural metadata fields to reduce token overhead.
The raw model response then passes through two repair stages before validation: a state-machine parser that resolves the vast majority of parse errors by correcting malformed string escapes, followed by an in-place repair loop that targets the remaining error positions exactly.
A final regex pass strips spurious thinking-channel tags that the model occasionally emits as a prefix.

\textbf{Hang detection and adaptive timeout.}
Long-context coding prompts occasionally caused the model to enter non-terminating generation loops, consistent with the cognitive-overload behaviour documented for SLMs in Section~\ref{sec:sota_memory}.
The backend addresses this at two levels.
At the output level, a tail-window scan inspects the end of every completed response and flags long runs of consecutive identical words as degenerate, discarding such outputs before they reach the parser.
At the request level, a health registry maintains rolling per-model statistics over mean response time and cumulative timeout rate; the effective timeout for each new request is set adaptively from this history, and models whose timeout rate exceeds a configured limit are marked unhealthy and assigned shorter timeouts to force earlier fallback.
An inactivity guard wraps all streaming calls and aborts if no token is received within a configurable window, preventing silent hangs from consuming the full timeout budget.

\subsection{Experimental Configuration}
\label{sec:exp_config}

All runs were executed on commodity workstations (AMD Ryzen~5 5600GE / Intel Core i7-8565U, 16\,GB RAM) running Windows~11 Pro.
Gemma~4~31B was accessed exclusively via Google AI Studio's hosted API; embeddings were generated with Google's hosted embedding model.
The baseline condition was produced by toggling the CBR-disable environment variable, which switches off retrieval and context injection while leaving quality gating and case retention active.
The retrieval thresholds used throughout the evaluation are an embedding-similarity threshold of~0.82 and a code-fingerprint threshold of~0.10 for the heuristic reuse-detection check (Section~\ref{sec:causal_reuse}); the deduplication threshold for G4 is~0.92.

%%%%%%%%%%%%%%%%%%%%%%%%%%%%%%%%%%%%%%%%%%%%%%%%%%%%%%%%%%%%%%%%%%%%%%%%%%%%%%%%
\section{Experimental Evaluation}
\label{sec:eval}

\subsection{Experimental Setup}

I evaluate CBR-augmented R\&D-Agent across two Kaggle competitions: \emph{NOMAD~2018 -- Predict Transparent Conductors} (RMSLE, lower is better) and \emph{Spaceship Titanic} (classification accuracy, higher is better).
The selection was constrained by the experimental window and hardware (commodity workstations without discrete GPUs); GPU-heavy vision or deep-learning competitions were excluded.
The two competitions span regression on a structured materials-science dataset (NOMAD) and binary classification on a generic tabular dataset (Spaceship Titanic), providing complementary CPU-feasible coverage.
Each competition is run under two conditions --- \textbf{baseline} (CBR off) and \textbf{within-comp} (CBR active, knowledge base seeded with cases from the same competition) --- with four random seeds each, yielding 16~runs in total.
All runs use the same backbone (Gemma~4~31B) and execute eight autonomous improvement loops.

\textbf{Scope: within-competition evaluation only.}
No cross-session condition is included; the preconditions and rationale are discussed in Section~\ref{sec:discussion}.

The remaining setup parameters are as follows.
All runs completed fully, with the exception of baseline seed~4 on Spaceship Titanic, which completed only seven usable loops due to a tooling crash in the final iteration.
Scores are reported as mean~$\pm$~standard deviation across seeds.
Effect sizes are reported as Cohen's~$d$; for completeness, two-sided Mann-Whitney U~$p$-values and bootstrapped 95\% confidence intervals (10\,000 resamples) are included but should be treated as descriptive rather than confirmatory: with $n=4$ per group the minimum achievable two-sided Mann-Whitney $p$-value is approximately~$0.03$, so the test has very limited power and all comparisons should be interpreted as exploratory.

The seed knowledge base used for all within-comp runs contains 60~successful cases drawn equally from both competitions, dominated by gradient-boosted tree models (XGBoost, LightGBM, and their ensembles), plus two recorded failure patterns.
The complete composition and per-run KB growth statistics appear in Appendix Tables~\ref{tab:a4_seed_kb} and~\ref{tab:a5_kb_stats}.

\textbf{Outlier handling.}
Score trajectories are computed as cumulative best per loop.
At each loop, a one-sided IQR test ($k=1.5$) flags values worse than $\text{Q3}+1.5\cdot\text{IQR}$ (lower-better) or $\text{Q1}-1.5\cdot\text{IQR}$ (higher-better), excluding only bad outliers while retaining beneficial ones.
Two events were identified: a within-comp run on NOMAD that stagnated visibly in the final two loops while all other within-comp runs converged to a tighter range (shown as a dashed line in Fig.~\ref{fig:score_trajectories}); and one baseline run on Spaceship Titanic with an anomalously low loop-0 accuracy (excluded from the loop-0 mean only; the run recovered by loop~1 and is otherwise included).

\textbf{Internal scores and public leaderboard correspondence.}
All metric values reported in this paper are internally measured validation scores produced by R\&D-Agent's feedback judge on the agent's local hold-out set.
Submission of one exemplary baseline run on Spaceship Titanic to the Kaggle public leaderboard revealed a systematic gap of approximately 0.012 between internal and public scores, attributable to local validation overfitting during iterative hypothesis refinement.
External validation results are discussed in Section~\ref{sec:external_validation}.

\subsection{Score Performance}

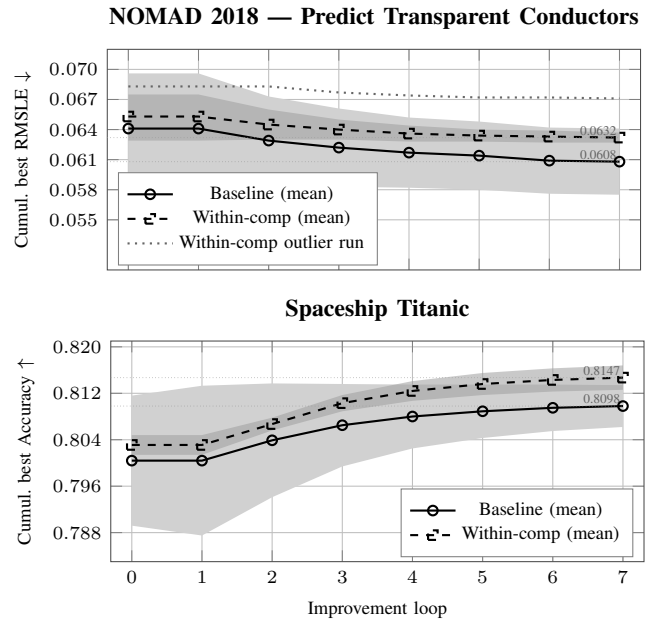
\begin{figure}[thpb]
\centering
\pgfplotsset{
  traj/.style={
    width=\columnwidth,
    height=4.5cm,
    xmin=-0.3, xmax=7.3,
    xtick={0,1,2,3,4,5,6,7},
    tick label style={font=\scriptsize},
    label style={font=\scriptsize},
    legend style={
      font=\scriptsize,
      draw=black!50,
      fill=white,
      fill opacity=1,          % fully opaque — covers any line behind it
      text opacity=1,
      inner sep=3pt,
      row sep=-1pt,
    },
    grid=both,
    grid style={line width=0.3pt, draw=black!15},
    major grid style={line width=0.4pt, draw=black!25},
    axis line style={black!70},
    clip=false,
  }
}

% ── PANEL A: NOMAD RMSLE ─────────────────────────────────────
\begin{tikzpicture}
\begin{axis}[
  traj,
  ylabel={Cumul.\ best RMSLE $\downarrow$},
  ymin=0.050, ymax=0.072,
  ytick={0.055,0.058,0.061,0.064,0.067,0.070},
  scaled y ticks=false,
  yticklabel style={
    /pgf/number format/fixed,
    /pgf/number format/precision=3,
    /pgf/number format/fixed zerofill=true,
  },
  xticklabels={,,},
  title style={font=\small\bfseries, yshift=-2pt},
  title={NOMAD 2018 --- Predict Transparent Conductors},
  % Legend south-west: below the SD bands in the early-loop empty area
  legend style={at={(0.02,0.02)}, anchor=south west},
]

% SD band Baseline
\addplot [name path=nb_hi, draw=none, forget plot] coordinates {
  (0,0.0696)(1,0.0696)(2,0.0673)(3,0.0661)(4,0.0652)(5,0.0648)(6,0.0642)(7,0.0641)};
\addplot [name path=nb_lo, draw=none, forget plot] coordinates {
  (0,0.0586)(1,0.0586)(2,0.0585)(3,0.0583)(4,0.0582)(5,0.0580)(6,0.0576)(7,0.0575)};
\addplot [black!18, opacity=1, forget plot]
  fill between[of=nb_hi and nb_lo];

% SD band Within-comp
\addplot [name path=nw_hi, draw=none, forget plot] coordinates {
  (0,0.0675)(1,0.0675)(2,0.0660)(3,0.0650)(4,0.0644)(5,0.0640)(6,0.0639)(7,0.0637)};
\addplot [name path=nw_lo, draw=none, forget plot] coordinates {
  (0,0.0629)(1,0.0629)(2,0.0630)(3,0.0630)(4,0.0628)(5,0.0628)(6,0.0627)(7,0.0627)};
\addplot [black!38, opacity=0.55, forget plot]
  fill between[of=nw_hi and nw_lo];

% Baseline mean — solid, circle markers
\addplot [black, thick, solid, mark=o, mark size=1.8pt,
          mark options={fill=white, draw=black}] coordinates {
  (0,0.0641)(1,0.0641)(2,0.0629)(3,0.0622)(4,0.0617)(5,0.0614)(6,0.0609)(7,0.0608)};
\addlegendentry{Baseline (mean)}

% Within-comp mean — dashed, square markers
\addplot [black, thick, dashed, mark=square, mark size=1.8pt,
          mark options={fill=black, draw=black}] coordinates {
  (0,0.0653)(1,0.0653)(2,0.0645)(3,0.0640)(4,0.0636)(5,0.0634)(6,0.0633)(7,0.0632)};
\addlegendentry{Within-comp (mean)}

% Outlier seed — dotted, no marker
\addplot [black!60, thick, dotted, mark=none] coordinates {
  (0,0.0683)(1,0.0683)(2,0.0683)(3,0.0677)(4,0.0674)(5,0.0672)(6,0.0672)(7,0.0671)};
\addlegendentry{Within-comp outlier run}

% Final value reference lines
\draw [black!40, very thin, densely dotted]
  (axis cs:-0.3,0.0608) -- (axis cs:7.0,0.0608)
  node [above left, font=\tiny, black!60, inner sep=1pt] {0.0608};
\draw [black!40, very thin, densely dotted]
  (axis cs:-0.3,0.0632) -- (axis cs:7.0,0.0632)
  node [above left, font=\tiny, black!60, inner sep=1pt] {0.0632};

\end{axis}
\end{tikzpicture}

\vspace{1pt}

% ── PANEL B: Spaceship Accuracy ──────────────────────────────
\begin{tikzpicture}
\begin{axis}[
  traj,
  xlabel={Improvement loop},
  ylabel={Cumul.\ best Accuracy $\uparrow$},
  ymin=0.783, ymax=0.821,
  % Fewer y-ticks — step 0.008 instead of 0.004
  ytick={0.788,0.796,0.804,0.812,0.820},
  scaled y ticks=false,
  yticklabel style={
    /pgf/number format/fixed,
    /pgf/number format/precision=3,
    /pgf/number format/fixed zerofill=true,
  },
  title style={font=\small\bfseries, yshift=-2pt},
  title={Spaceship Titanic},
  % Legend south-east: once lines have risen the bottom-right is clear
  legend style={at={(0.98,0.02)}, anchor=south east},
]

% SD band Baseline
\addplot [name path=sb_hi, draw=none, forget plot] coordinates {
  (0,0.8116)(1,0.8133)(2,0.8137)(3,0.8136)(4,0.8135)(5,0.8135)(6,0.8135)(7,0.8134)};
\addplot [name path=sb_lo, draw=none, forget plot] coordinates {
  (0,0.7892)(1,0.7875)(2,0.7941)(3,0.7994)(4,0.8025)(5,0.8043)(6,0.8055)(7,0.8062)};
\addplot [black!18, opacity=1, forget plot]
  fill between[of=sb_hi and sb_lo];

% SD band Within-comp
\addplot [name path=sw_hi, draw=none, forget plot] coordinates {
  (0,0.8048)(1,0.8048)(2,0.8078)(3,0.8117)(4,0.8141)(5,0.8155)(6,0.8163)(7,0.8168)};
\addplot [name path=sw_lo, draw=none, forget plot] coordinates {
  (0,0.8014)(1,0.8014)(2,0.8056)(3,0.8089)(4,0.8107)(5,0.8117)(6,0.8123)(7,0.8126)};
\addplot [black!38, opacity=0.55, forget plot]
  fill between[of=sw_hi and sw_lo];

% Baseline mean — solid, circle
\addplot [black, thick, solid, mark=o, mark size=1.8pt,
          mark options={fill=white, draw=black}] coordinates {
  (0,0.8004)(1,0.8004)(2,0.8039)(3,0.8065)(4,0.8080)(5,0.8089)(6,0.8095)(7,0.8098)};
\addlegendentry{Baseline (mean)}

% Within-comp mean — dashed, square
\addplot [black, thick, dashed, mark=square, mark size=1.8pt,
          mark options={fill=black, draw=black}] coordinates {
  (0,0.8031)(1,0.8031)(2,0.8067)(3,0.8103)(4,0.8124)(5,0.8136)(6,0.8143)(7,0.8147)};
\addlegendentry{Within-comp (mean)}

% Final value reference lines
\draw [black!40, very thin, densely dotted]
  (axis cs:-0.3,0.8098) -- (axis cs:7.0,0.8098)
  node [above left, font=\tiny, black!60, inner sep=1pt] {0.8098};
\draw [black!40, very thin, densely dotted]
  (axis cs:-0.3,0.8147) -- (axis cs:7.0,0.8147)
  node [above left, font=\tiny, black!60, inner sep=1pt] {0.8147};

\end{axis}
\end{tikzpicture}

\caption{Cumulative-best score trajectories across eight autonomous improvement loops.
\textit{Top:} NOMAD (RMSLE, lower is better). \textit{Bottom:} Spaceship Titanic (accuracy, higher is better).
Solid line with open circles = Baseline mean; dashed line with filled squares = Within-comp mean.
Shaded bands = $\pm$1\,SD (outlier loop values excluded from the band).
Dotted line = one within-comp run on NOMAD that stagnated in the final two loops and was excluded from the band (but retained in all other analyses).
$n = 4$ seeds per condition.}
\label{fig:score_trajectories}
\end{figure}

Fig.~\ref{fig:score_trajectories} shows the cumulative-best score trajectories for both competitions.
Shaded bands indicate $\pm1$ standard deviation across seeds (excluding flagged outlier values); dotted lines denote runs containing at least one excluded loop.

\ifx\tikzpicture\undefined\endinput\fi
% ============================================================
% Figure 2 — Score Trajectories  (BLACK & WHITE VERSION)
% Baseline:    solid line,  circle markers  (○)
% Within-comp: dashed line, square markers  (□)
% Outlier:     dotted line, no markers
% SD bands:    grey fills
% Legend positions chosen to avoid data overlap:
%   NOMAD       → south-west (empty space below bands at early loops)
%   Spaceship   → south-east (empty space when lines have climbed high)
% ============================================================

Table~\ref{tab:1_score_performance} summarises the key performance metrics per group.

\begin{table}[t]
  \centering
  \setlength{\tabcolsep}{2pt}
  \scriptsize
  \caption{Score performance per condition group.
    NOMAD: RMSLE\,$\downarrow$; Spaceship (Space.): Accuracy\,$\uparrow$.
    \textbf{Best} = best cumul.\ metric (8~loops), mean\,$\pm$\,SD.
    \textbf{L0} = loop-0 score.
    \textbf{$\Delta$abs} = abs.\ SOTA gain (mean\,$\pm$\,SD).
    \textbf{$\Delta$\%} = rel.\ SOTA gain.
    \textbf{L$_1$} = loops to first improvement (mean).
    $^\dagger$NOMAD base.\ $n{=}3$ (seed~2 best in loop~0); $n{=}4$ otherwise.}
  \label{tab:1_score_performance}
  \begin{tabular}{@{}llccccc@{}}
    \toprule
    \textbf{Comp.} & \textbf{Cond.} &
    \textbf{Best} & \textbf{L0} &
    \textbf{$\Delta$abs} & \textbf{$\Delta$\%} & \textbf{L$_1$} \\
    \midrule
    NOMAD  & Base.   & $0.0608{\pm}0.0038$ & 0.0641 & $0.0044{\pm}0.0025$$^\dagger$ & 6.50\%$^\dagger$ & 2.00 \\
    NOMAD  & W-comp  & $0.0632{\pm}0.0027$ & 0.0653 & $0.0009{\pm}0.0009$           & 1.41\%           & 2.25 \\
    \midrule
    Space. & Base.   & $0.8098{\pm}0.0042$ & 0.8004 & $0.0095{\pm}0.0131$           & 1.21\%           & 1.25 \\
    Space. & W-comp  & $0.8147{\pm}0.0024$ & 0.8031 & $0.0116{\pm}0.0013$           & 1.45\%           & 1.25 \\
    \bottomrule
  \end{tabular}
\end{table}

On NOMAD, the baseline achieves a lower (better) mean RMSLE than within-comp (0.0608 vs.\ 0.0632, $d=-0.75$, $p=0.20$).
The within-comp condition shows a substantially smaller absolute SOTA gain across the run (0.0009 vs.\ 0.0044) and a smaller relative gain (1.4\% vs.\ 6.5\%), indicating that the baseline explores the performance landscape more broadly over eight loops.
On Spaceship Titanic the picture reverses: within-comp reaches a higher mean accuracy (0.8147 vs.\ 0.8098, $d=-1.41$, $p=0.11$) and a marginally larger absolute SOTA gain.
Notably, within-comp also exhibits substantially lower variance on Spaceship Titanic in both best metric and SOTA gain (see Table~\ref{tab:1_score_performance}), suggesting more consistent learning behaviour under CBR.
Per-run scores and statistical comparisons are in Appendix Tables~\ref{tab:a1_per_run_perf} and~\ref{tab:a3_stat_tests}.

\subsection{Efficiency and Coding Effort}

Table~\ref{tab:2_efficiency} reports token consumption, wall-clock duration, and internal iteration counts per run.

\begin{table}[t]
  \centering
  \setlength{\tabcolsep}{3pt}
  \footnotesize
  \caption{Efficiency per condition group (means).
    \textbf{Pr\,(k)} / \textbf{Co\,(k)} = prompt / completion tokens (k).
    \textbf{Dur} = wall-clock duration (min).
    \textbf{RE/L} = run evo-loops per improvement loop.
    \textbf{CE/L} = CoSTEER iterations per loop.
    \textbf{Ret} = retention rate.
    \textbf{Bold}: only result with $p{<}0.05$ ($p{=}0.029$, MW). $n{=}4$.}
  \label{tab:2_efficiency}
  \begin{tabular}{llcccccc}
    \toprule
    \textbf{Comp.} & \textbf{Cond.} &
    \textbf{Pr\,(k)} & \textbf{Co\,(k)} & \textbf{Dur} &
    \textbf{RE/L} & \textbf{CE/L} & \textbf{Ret} \\
    & & \textit{\scriptsize m$\pm$SD} & \textit{\scriptsize m} &
    \textit{\scriptsize m} & \textit{\scriptsize m} &
    \textit{\scriptsize m} & \textit{\scriptsize m} \\
    \midrule
    NOMAD & Base.  & $877\pm226$          & 150 & 160 & 2.38 & 1.28 & 0.43 \\
    NOMAD & W-comp & $913\pm30$           & 144 & 160 & 2.50 & 1.19 & 0.44 \\
    \midrule
    Space.& Base.  & $625\pm87$           & 127 & 141 & 2.39 & 1.31 & 0.42 \\
    Space.& W-comp & $\mathbf{929\pm229}$ & 132 & 138 & 2.25 & 1.31 & 0.56 \\
    \bottomrule
  \end{tabular}
\end{table}

On NOMAD, token consumption, wall-clock duration, and coding effort are essentially indistinguishable across the two conditions.
On Spaceship Titanic, within-comp consumes substantially more prompt tokens than baseline ($d=-1.76$, $p=0.029$).
Completion tokens differ only marginally.
Despite the higher prompt-token load, wall-clock duration is slightly shorter for within-comp on Spaceship Titanic (138 vs.\ 141~min), suggesting that the additional context may reduce the number of internal revision cycles needed, thereby partially offsetting the latency cost.

The retention rate shows a directionally consistent advantage for within-comp across both competitions.
The effect is notably larger on Spaceship Titanic, where the within-comp condition accepted on average 4.5 experiments per run against 3.25 for the baseline, indicating that CBR-informed hypotheses are more consistently judged as acceptable by the feedback model.
The number of running and coding evo-loops per improvement loop is comparable across all conditions, confirming that CBR does not systematically alter the difficulty of the code-generation phase.
Full per-run token and timing details are provided in Appendix Table~\ref{tab:a2_per_run_tokens}.

\subsection{Knowledge Base Growth and Reuse}

Each within-comp run grew the KB from the 60-case seed to 67--68 cases over eight loops, with 2--3 new failure patterns per run.
Retrieval engagement is comparable across the two competitions: between six and ten distinct cases are actively retrieved at least once per run, with 27--31 total retrieval attempts per run of which 11--15 meet both similarity thresholds of the heuristic reuse-detection check (Section~\ref{sec:causal_reuse}).
The complete per-run KB statistics appear in Appendix Table~\ref{tab:a5_kb_stats}.

\subsection{Heuristic Reuse Detection}

\begin{figure}[thpb]
\centering
\begin{tikzpicture}
\begin{axis}[
  width=\columnwidth,
  height=6.2cm,
  xmin=0.818, xmax=0.962,
  ymin=-0.02, ymax=0.82,
  xtick={0.82,0.84,0.86,0.88,0.90,0.92,0.94,0.96},
  ytick={0.0,0.1,0.2,0.3,0.4,0.5,0.6,0.7,0.8},
  scaled x ticks=false,
  xticklabel style={/pgf/number format/fixed,
                    /pgf/number format/precision=2},
  tick label style={font=\scriptsize},
  label style={font=\scriptsize},
  xlabel={Embedding cosine similarity},
  ylabel={Code fingerprint similarity},
  grid=both,
  grid style={line width=0.3pt, draw=black!15},
  major grid style={line width=0.4pt, draw=black!25},
  axis line style={black!70},
  legend style={
    font=\scriptsize,
    at={(0.02,0.98)},
    anchor=north west,
    draw=black!50,
    fill=white,
    inner sep=3pt,
    row sep=0pt,
  },
  clip=true,
]

% ── NOMAD: open circles ──────────────────────────────────────
\addplot [
  only marks,
  mark=o,
  mark size=1.6pt,
  mark options={fill=white, draw=black, line width=0.5pt},
  black,
] coordinates {
  (0.8995,0.100)(0.8925,0.100)(0.9011,0.343)(0.9108,0.100)
  (0.8914,0.100)(0.8914,0.593)(0.9114,0.226)(0.9024,0.290)
  (0.8888,0.100)(0.9000,0.157)(0.8889,0.299)(0.8889,0.100)
  (0.8967,0.357)(0.8730,0.145)(0.8750,0.601)(0.8667,0.102)
  (0.9421,0.185)(0.8885,0.165)(0.8499,0.102)(0.9194,0.119)
  (0.8438,0.147)(0.8967,0.314)(0.8220,0.236)(0.8399,0.102)
  (0.8963,0.234)(0.9163,0.363)(0.8860,0.117)(0.9298,0.568)
  (0.9298,0.423)(0.9298,0.197)(0.8805,0.423)(0.8985,0.688)
  (0.9203,0.334)(0.9298,0.694)(0.8927,0.117)(0.9027,0.527)
  (0.8714,0.362)(0.8683,0.275)(0.9150,0.363)(0.9298,0.680)
  (0.8782,0.101)(0.9013,0.200)(0.9298,0.101)(0.8663,0.307)
  (0.8547,0.302)(0.8669,0.262)(0.9236,0.225)(0.9002,0.101)
  (0.8658,0.196)(0.9075,0.208)(0.8909,0.137)(0.9257,0.236)
  (0.8589,0.494)
};
\addlegendentry{NOMAD ($n{=}53$)}

% ── Spaceship: filled squares ────────────────────────────────
\addplot [
  only marks,
  mark=square*,
  mark size=1.5pt,
  mark options={fill=black, draw=black},
  black,
] coordinates {
  (0.8968,0.113)(0.8643,0.170)(0.8659,0.436)(0.8596,0.114)
  (0.8245,0.343)(0.8605,0.462)(0.8621,0.161)(0.9078,0.210)
  (0.8573,0.287)(0.8667,0.217)(0.8603,0.113)(0.8388,0.283)
  (0.8827,0.153)(0.8753,0.448)(0.8487,0.163)(0.9005,0.404)
  (0.8516,0.419)(0.8612,0.255)(0.8851,0.401)(0.8422,0.411)
  (0.8728,0.250)(0.8954,0.663)(0.8342,0.440)(0.8719,0.173)
  (0.8735,0.469)(0.8844,0.208)(0.8420,0.490)(0.8403,0.549)
  (0.8790,0.604)(0.9128,0.101)(0.8946,0.269)(0.9081,0.757)
  (0.9298,0.439)(0.8729,0.155)(0.8561,0.115)(0.8516,0.411)
  (0.8541,0.276)(0.8785,0.455)(0.8923,0.409)(0.8901,0.305)
  (0.9083,0.157)(0.8814,0.606)(0.8613,0.191)(0.8926,0.191)
  (0.8771,0.232)(0.8421,0.461)(0.8762,0.461)(0.8812,0.461)
  (0.8508,0.762)(0.8757,0.470)(0.8734,0.559)(0.8446,0.531)
  (0.8806,0.483)(0.8855,0.330)(0.8980,0.617)
};
\addlegendentry{Spaceship ($n{=}55$)}

% ── Overall mean cross ───────────────────────────────────────
\addplot [
  only marks, mark=+,
  mark size=4.5pt,
  mark options={line width=1.4pt, draw=black},
  black,
] coordinates { (0.882,0.305) };
\addlegendentry{Overall mean}

% ── Threshold lines ──────────────────────────────────────────
% Vertical: emb threshold = 0.82
\draw [black!70, dashed, line width=0.7pt]
  (axis cs:0.820,-0.02) -- (axis cs:0.820,0.82);

% Horizontal: fp threshold = 0.10
\draw [black!70, dashed, line width=0.7pt]
  (axis cs:0.818,0.10) -- (axis cs:0.962,0.10);

% Labels
\node [font=\tiny, text=black!65, rotate=90, anchor=south]
  at (axis cs:0.8185,0.48) {emb.\ thr.\,$=0.82$};
\node [font=\tiny, text=black!65, anchor=south west]
  at (axis cs:0.819,0.103) {fp thr.\,$=0.10$};

% Mean annotation
\node [font=\tiny, anchor=south west, xshift=3pt, yshift=1pt]
  at (axis cs:0.882,0.305) {$(0.882,\,0.305)$};

\end{axis}
\end{tikzpicture}

\caption{Heuristic reuse-detection scatter: embedding cosine similarity ($x$-axis)
vs.\ code-fingerprint similarity ($y$-axis) for all 108 logged CBR
retrieval events (NOMAD within-comp: open circles, $n{=}53$;
Spaceship Titanic within-comp: filled squares, $n{=}55$).
Cross = overall mean $(0.882,\,0.305)$.
Dashed lines: vertical = embedding retrieval threshold (0.82);
horizontal = fingerprint threshold (0.10).
All 108 events exceeded both thresholds; no heuristic reuse-detection check returned a negative outcome.}
\label{fig:causality_scatter}
\end{figure}
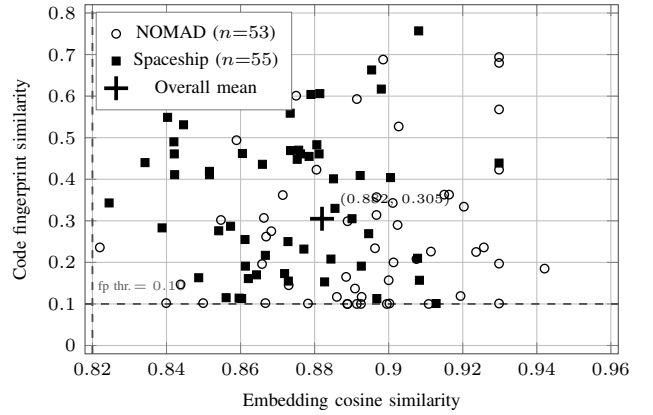

Fig.~\ref{fig:causality_scatter} shows the embedding similarity versus code-fingerprint similarity for all 108 logged CBR retrieval events from the within-comp runs.

\ifx\tikzpicture\undefined\endinput\fi
% ============================================================
% Figure 3 — Heuristic Reuse Detection Scatter  (BLACK & WHITE VERSION)
% NOMAD:     open circles  (○)
% Spaceship: filled squares (■)
% Mean:      cross (+)
% x-axis range: 0.820–0.960  (all real data in this range)
% ============================================================

All 108 retrieval events met both similarity thresholds; no heuristic reuse-detection check returned a negative outcome.
Mean embedding similarity was 0.882 (SD~$0.031$), well above the retrieval threshold, indicating that retrieved cases are consistently semantically close to the current problem context.
Code-fingerprint similarity was considerably more variable (mean~$0.305$, SD~$0.176$), with values spread from the structural-similarity floor up to roughly~0.76; all events met or exceeded the fingerprint threshold.
This wide spread above the floor indicates that the degree of structural alignment varies substantially with the specificity of the retrieved case and the generative behaviour of the LLM.
Per-run mean embedding similarity remained tight across all eight within-comp runs, while per-run mean fingerprint similarity varied roughly two-fold (see Appendix Table~\ref{tab:a6_causal_reuse}).
This asymmetry --- high and stable embedding similarity alongside more variable fingerprint similarity --- is consistent with CBR influencing hypothesis generation at a structural level rather than enabling direct code copying, though this interpretation cannot be established without controlled retrieval ablations.

\subsection{Exemplary Baseline: External Validation}
\label{sec:external_validation}

To complement the internal evaluation, I validated one exemplary baseline run against two external references.

\textbf{Kaggle public leaderboard (Spaceship Titanic).}
All eight per-loop experiment outputs of one baseline run were submitted to the Kaggle public leaderboard (2\,240 teams as of 18~May~2026).
All submissions exceeded the leaderboard median; the peak public score landed exactly at that median.

\textbf{MLE-bench grading (NOMAD~2018).}
The same exemplary baseline run was evaluated against the MLE-bench grading pipeline on the held-out test split.
Seven of the eight loops achieved a silver medal under the MLE-bench thresholds for this competition; the single non-medal loop was loop~0, before any iterative refinement.
The SOTA score recorded by the grader corresponds to a silver-medal result.

%%%%%%%%%%%%%%%%%%%%%%%%%%%%%%%%%%%%%%%%%%%%%%%%%%%%%%%%%%%%%%%%%%%%%%%%%%%%%%%%
\section{Discussion}
\label{sec:discussion}

\subsection{CBR Effect on Final Performance}

The results present a competition-dependent picture.
On NOMAD, CBR does not improve --- and by the best-metric measure produces a slightly worse --- mean RMSLE compared with the baseline ($d=-0.75$), though the 95\% confidence interval spans zero and the effect does not reach statistical significance at the available sample size.
One possible explanation for the smaller SOTA gain under within-comp on NOMAD is that retrieved cases cluster around a particular RMSLE level, causing the agent to generate hypotheses anchored to that level rather than exploring more aggressively.
This pattern is broadly consistent with reduced exploration, but it should be noted that this interpretation is post-hoc and based solely on observational trajectory data: no experiment withheld or varied the KB composition to test whether anchoring to prior scores is the causal mechanism.
Verifying this claim would require, for example, comparing runs whose seed KBs differ in performance spread, which is deferred to future work.
The one within-comp NOMAD run that stagnated most severely (flagged as an outlier in Section~\ref{sec:eval}) is broadly consistent with this interpretation, but a single run does not constitute evidence for the proposed mechanism.

On Spaceship Titanic the picture reverses: within-comp shows a directional advantage in both final score ($d=-1.41$) and total SOTA gain.
Here the KB does not appear to limit final score, though the mechanism behind this difference remains unclear from the available data.
One observable difference is that classification accuracy on this dataset exhibits a natural distributional ceiling, which may structurally reduce anchoring effects compared to a regression task; however, this is a post-hoc observation rather than an experimentally tested claim.
The substantially lower variance of within-comp scores on Spaceship Titanic is an observable result; whether it reflects more consistent exploration or other factors such as task difficulty is not determinable from the current data.

\subsection{CBR Effect on Learning Dynamics}

Despite the mixed picture for final scores, both competitions show a directional advantage for within-comp in retention rate: CBR-informed hypotheses are more frequently judged as acceptable by the feedback model, particularly on Spaceship Titanic.
This indicates that retrieved cases improve local hypothesis quality --- the agent generates more immediately executable and acceptable code --- even when this does not translate into a superior global optimum.
First-improvement speed is essentially identical across conditions, confirming that CBR does not accelerate the initial learning phase.

The trajectory plots reveal a qualitative difference in exploration dynamics.
Baseline runs tend to show larger individual score jumps alongside higher inter-run variance, consistent with broader, less-constrained exploration.
Within-comp runs show smaller, more consistent per-loop improvements and tighter variance bands.
The trajectory patterns suggest that CBR concentrates the search around regions of the performance landscape that have previously proven useful, trading variance reduction for potentially reduced exploratory breadth.
Whether this consistency-exploration trade-off is beneficial depends on the available loop budget and competition structure; within the eight-loop budget of these experiments, the two conditions yield broadly comparable results, with the advantage shifting between competitions.
Labelling this as an ``exploitative strategy'' would be an over-interpretation of the available data: demonstrating that CBR systematically suppresses exploration would require an experiment that controls loop budget and KB composition while varying retrieval breadth, which was not performed here.

\subsection{Token Efficiency and Cost}

CBR retrieval adds a meaningful prompt-token overhead ($+49\%$ on Spaceship Titanic), reflecting injected case context in every proposal prompt; this does not translate proportionally into wall-clock time, suggesting the context reduces internal revision cycles.
On NOMAD, where baseline prompt lengths are already large and highly variable, the CBR overhead is masked by baseline variance and shows no detectable difference.
The coding effort (evo-loops per improvement loop) is comparable across all conditions, confirming that CBR does not affect the difficulty of the code-generation phase itself.

\subsection{Heuristic Reuse Detection}

The heuristic reuse-detection analysis shows that CBR retrieval consistently identifies semantically relevant cases while the structural proximity of the generated code to retrieved examples varies considerably.
All events cleared the structural floor; the wide spread above it reflects varying structural alignment depending on case specificity and LLM generative behaviour --- consistent with conceptual guidance rather than verbatim copying.
The uniformly high embedding similarity and absence of negative outcomes are expected given the within-comp KB contains exclusively same-competition cases, maximising retrieval relevance.

\subsection{External Validation in Context}

On Spaceship Titanic, all eight per-loop submissions exceeded the public leaderboard median; the best loop reached exactly that median, placing the agent's output in the upper half of the 2\,240-team field within eight fully autonomous loops.

On NOMAD~2018, the MLE-bench grading pipeline awarded a silver medal to seven of the eight loops of the same baseline run.
Under the MLE-bench medal rules for this competition's leaderboard size, this corresponds to a top-5--6\% placement on the held-out leaderboard.
This is a strong result for a locally deployable SLM operating without human guidance, particularly on a domain-specific materials-science task requiring structural feature engineering from crystallographic geometry files.

Both results derive from a single exemplary run (not the full four-seed protocol) and the MLE-bench split differs from the original Kaggle test set, but they provide external anchoring for the internal evaluation.

\subsection{Limitations and Outlook}

Several limitations constrain the generalisability of these results.
First, four seeds per group means no Mann-Whitney test can reach $p<0.05$ under the standard two-sided threshold; all effect sizes should be treated as preliminary estimates pending replication with larger samples.
Second, only two competitions are evaluated, both in the tabular machine learning domain on Kaggle; results may not generalise to other problem domains, knowledge-base compositions, or underlying LLMs.
Third, the NOMAD SOTA gain metric is computed over only three baseline runs, because one seed achieved its best score in loop~0 (rendering an absolute gain undefined), which reduces statistical power for that specific metric.
Fourth, the heuristic reuse-detection mechanism does not constitute causal analysis in the formal sense.
The three-condition check --- embedding similarity, code-fingerprint overlap, injection provenance --- provides a plausible proxy for knowledge transfer, but cannot rule out coincidental structural similarity or independent convergence to the same solution.
Establishing whether retrieved cases causally influence agent behaviour requires controlled experiments such as retrieval ablations (withholding a specific case and comparing downstream code), prompt-masking studies, or token-attribution analysis, none of which were performed here.
The reuse statistics reported in Section~\ref{sec:eval} should therefore be interpreted as descriptive indicators of co-occurrence rather than as evidence of causal influence.

Fifth, and most fundamentally, no cross-session condition was evaluated in this work.
This is not merely a gap in experimental coverage --- it is a deliberate consequence of the current state of the system.
Three conditions must be met before cross-session learning can be evaluated in a principled way: (a)~the KB must contain a sufficient number of high-quality cases across \emph{multiple} domains, not just two; (b)~the total number of trial runs executed must be large enough to produce a statistically stable retrieval behaviour across domain boundaries; and (c)~the range of competitions covered must be broad enough to distinguish genuine cross-domain transfer from coincidental similarity between two structurally similar datasets.
None of these conditions is currently met.
The within-competition evaluation thus validates architectural soundness and retrieval quality, but does not yet test the system's core value proposition of persistent, cross-session, cross-domain knowledge transfer.

The path towards that test is clear and constitutes the primary agenda for follow-on work.
\textbf{Step 1} is to extend the system so that it can run reliably across the full MLE-Bench suite of 75 competitions, covering the range of problem types (tabular, vision, NLP, time-series, and domain-specific engineering tasks) that the benchmark includes.
\textbf{Step 2} is to accumulate a sufficiently large and diverse KB by running the CBR agent across this broader competition set --- generating the cross-competition case diversity needed for a meaningful retrieval signal.
\textbf{Step 3} is to evaluate cross-session learning rigorously: a held-out competition is presented to the agent equipped with a KB built from prior, disjoint competitions, and retrieval quality, retention rate, and final performance are compared against both a no-CBR baseline and a within-competition oracle condition.
This three-step roadmap will allow the research question that motivates the entire architecture --- \emph{can structured, persistent case memory substitute for parameter scaling in an autonomous data-science agent?} --- to be answered with the rigour it deserves.

\textbf{Long-term deployment vision.}
Beyond benchmark performance, the deeper motivation for this line of work is a specific deployment scenario that current frontier-cloud-first systems do not address: small and medium-sized engineering teams that need autonomous data-science support but cannot or do not want to depend on external cloud APIs for reasons of cost, data privacy, infrastructure complexity, or the desire to keep their analytical pipelines fully auditable and under their own control.
The vision is a system that a small engineering team can install on their own servers, run without a cloud subscription or specialised ML-infrastructure expertise, inspect and extend by reading the source code, and trust because its memory is stored in plain, human-readable files.
This preprint establishes that such a system can be built and works; the follow-on roadmap above is the path from ``it works'' to ``it is competitive''.

%%%%%%%%%%%%%%%%%%%%%%%%%%%%%%%%%%%%%%%%%%%%%%%%%%%%%%%%%%%%%%%%%%%%%%%%%%%%%%%%
\section{Conclusion}
\label{sec:conclusion}

\textbf{This paper presented CBR-augmented R\&D-Agent: a persistent, quality-controlled case-based memory layer integrated into a locally runnable 31\,B-parameter agent backbone via a surgical three-method subclass. The Gemma~4 backend shows that an SLM can serve as a viable autonomous agent backbone through targeted engineering adaptations.\\[6pt]
Empirically, CBR improves consistency and retention rate across both test competitions.
On Spaceship Titanic it achieves a directional accuracy advantage with markedly lower variance; on NOMAD the baseline's broader exploration yields a larger absolute SOTA gain.
The heuristic reuse-detection analysis shows that the system consistently retrieves semantically relevant cases; patterns in structural similarity are consistent with conceptual guidance rather than code copying, though establishing this directionally requires controlled ablation studies deferred to future work.\\[6pt]
These results establish a first empirical data point for an SLM-based CBR-augmented autonomous data-science agent and point toward a design space that prioritises local deployability, transparency, and maintainability alongside benchmark performance.
The architecture is explicitly designed for cross-session, cross-domain reuse; extending evaluation to the full MLE-Bench suite is the logical next step.
As a preprint, this paper records a first implementation step toward a locally deployable, transparent agent for small engineering teams that accumulates and reuses experience on its own hardware --- the follow-on roadmap in Section~\ref{sec:discussion} constitutes the path to the full realisation.}

\section*{Acknowledgment}

The author thanks the R\&D-Agent team at Microsoft Research for open-sourcing their framework, and Google DeepMind for making Gemma~4 available through Google AI Studio. Parts of this manuscript were drafted and refined with the assistance
of Claude (Anthropic), a large language model used as an AI writing aid.
The source code of the CBR-augmented R\&D-Agent is publicly available at \url{https://github.com/stofe94/cbr-rd-agent}.

%%%%%%%%%%%%%%%%%%%%%%%%%%%%%%%%%%%%%%%%%%%%%%%%%%%%%%%%%%%%%%%%%%%%%%%%%%%%%%%%
\addtolength{\textheight}{-3cm}   % last-page column-balance adjustment
\bibliographystyle{IEEEtran}
\bibliography{bibliography}

%%%%%%%%%%%%%%%%%%%%%%%%%%%%%%%%%%%%%%%%%%%%%%%%%%%%%%%%%%%%%%%%%%%%%%%%%%%%%%%%
\clearpage
\appendix

\begin{table*}[t]
  \centering
  \small
  \caption{\textbf{Appendix~A: Per-Run Performance Metrics.} \textbf{Comp.}\ = competition. \textbf{Cond.}\ = condition. \textbf{S} = seed. \textbf{N} = loops completed. \textbf{Best} = best cumulative metric. \textbf{L0} = loop-0 metric. \textbf{$\Delta$SOTA} = absolute SOTA gain (direction-corrected); {\normalfont\textit{n/a}} = best score reached in loop~0. \textbf{$\Delta$\%} = relative SOTA gain. \textbf{L$_1$} = loops to first improvement. \textbf{Ret.}\ = retention rate. NOMAD: RMSLE $\downarrow$; Spaceship: Accuracy $\uparrow$.}
  \label{tab:a1_per_run_perf}
  \begin{tabular}{llllrrrrrrr}
    \toprule
    \textbf{Run ID} & \textbf{Comp.} & \textbf{Cond.} & \textbf{S} & \textbf{N} & \textbf{Best} & \textbf{L0} & \textbf{$\Delta$SOTA} & \textbf{$\Delta$\%} & \textbf{L$_1$} & \textbf{Ret.} \\
    \midrule
    \texttt{phase1\_A\_baseline\_seed1} & NOMAD & Baseline    & 1 & 8 & 0.0658 & 0.0710 & 0.0052 & 7.26\% & 1 & 0.625 \\
    \texttt{phase1\_A\_baseline\_seed2} & NOMAD & Baseline    & 2 & 8 & 0.0568 & 0.0568 & n/a    & n/a    & 1 & 0.125 \\
    \texttt{phase1\_A\_baseline\_seed3} & NOMAD & Baseline    & 3 & 8 & 0.0609 & 0.0674 & 0.0065 & 9.66\% & 1 & 0.714 \\
    \texttt{phase1\_A\_baseline\_seed4} & NOMAD & Baseline    & 4 & 8 & 0.0596 & 0.0612 & 0.0016 & 2.59\% & 5 & 0.250 \\
    \texttt{phase2\_A\_within\_seed1}   & NOMAD & Within-comp & 1 & 8 & 0.0671 & 0.0683 & 0.0012 & 1.71\% & 2 & 0.625 \\
    \texttt{phase2\_A\_within\_seed2}   & NOMAD & Within-comp & 2 & 8 & 0.0625 & 0.0646 & 0.0021 & 3.22\% & 1 & 0.500 \\
    \texttt{phase2\_A\_within\_seed3}   & NOMAD & Within-comp & 3 & 8 & 0.0612 & 0.0613 & 0.0002 & 0.26\% & 5 & 0.250 \\
    \texttt{phase2\_A\_within\_seed4}   & NOMAD & Within-comp & 4 & 8 & 0.0621 & 0.0668 & 0.0003 & 0.45\% & 1 & 0.375 \\
    \midrule
    \texttt{phase1\_B\_baseline\_seed1} & Spaceship & Baseline    & 1 & 8 & 0.8090 & 0.8055 & 0.0036 & 0.44\% & 1 & 0.375 \\
    \texttt{phase1\_B\_baseline\_seed2} & Spaceship & Baseline    & 2 & 8 & 0.8133 & 0.8092 & 0.0041 & 0.51\% & 2 & 0.750 \\
    \texttt{phase1\_B\_baseline\_seed3} & Spaceship & Baseline    & 3 & 8 & 0.8128 & 0.8116 & 0.0013 & 0.16\% & 1 & 0.250 \\
    \texttt{phase1\_B\_baseline\_seed4} & Spaceship & Baseline    & 4 & 7 & 0.8042 & 0.7752 & 0.0290 & 3.74\% & 1 & 0.286 \\
    \texttt{phase2\_B\_within\_seed1}   & Spaceship & Within-comp & 1 & 8 & 0.8161 & 0.8027 & 0.0133 & 1.66\% & 1 & 0.500 \\
    \texttt{phase2\_B\_within\_seed2}   & Spaceship & Within-comp & 2 & 8 & 0.8143 & 0.8026 & 0.0117 & 1.46\% & 1 & 0.625 \\
    \texttt{phase2\_B\_within\_seed3}   & Spaceship & Within-comp & 3 & 8 & 0.8170 & 0.8058 & 0.0112 & 1.38\% & 2 & 0.375 \\
    \texttt{phase2\_B\_within\_seed4}   & Spaceship & Within-comp & 4 & 8 & 0.8115 & 0.8012 & 0.0102 & 1.28\% & 1 & 0.750 \\
    \bottomrule
  \end{tabular}
\end{table*}

\begin{table*}[t]
  \centering
  \small
  \caption{\textbf{Appendix~B: Per-Run Token Consumption and Timing.} \textbf{Prompt\,(k)} / \textbf{Compl.\,(k)} = prompt / completion tokens in thousands (rounded). \textbf{Dur.}\ = total wall-clock duration in minutes. \textbf{Run\,evo/L} = mean running evo-loops per improvement loop (internal code-execution attempts). \textbf{Cod\,evo/L} = mean CoSTEER debug iterations per improvement loop.}
  \label{tab:a2_per_run_tokens}
  \begin{tabular}{lllrrrrr}
    \toprule
    \textbf{Run ID} & \textbf{Comp.} & \textbf{Cond.} & \textbf{Prompt\,(k)} & \textbf{Compl.\,(k)} & \textbf{Dur.\,(min)} & \textbf{Run\,evo/L} & \textbf{Cod\,evo/L} \\
    \midrule
    \texttt{phase1\_A\_baseline\_seed1} & NOMAD & Baseline    &  847 & 160 & 179.9 & 2.62 & 1.12 \\
    \texttt{phase1\_A\_baseline\_seed2} & NOMAD & Baseline    &  609 & 124 & 117.6 & 2.62 & 1.00 \\
    \texttt{phase1\_A\_baseline\_seed3} & NOMAD & Baseline    &  889 & 172 & 181.3 & 1.88 & 1.62 \\
    \texttt{phase1\_A\_baseline\_seed4} & NOMAD & Baseline    & 1162 & 144 & 160.4 & 2.38 & 1.38 \\
    \texttt{phase2\_A\_within\_seed1}   & NOMAD & Within-comp &  938 & 150 & 182.7 & 2.50 & 1.25 \\
    \texttt{phase2\_A\_within\_seed2}   & NOMAD & Within-comp &  937 & 147 & 157.3 & 2.62 & 1.12 \\
    \texttt{phase2\_A\_within\_seed3}   & NOMAD & Within-comp &  878 & 137 & 145.7 & 2.50 & 1.25 \\
    \texttt{phase2\_A\_within\_seed4}   & NOMAD & Within-comp &  898 & 142 & 152.3 & 2.38 & 1.12 \\
    \midrule
    \texttt{phase1\_B\_baseline\_seed1} & Spaceship & Baseline    &  616 & 131 & 141.6 & 2.00 & 1.62 \\
    \texttt{phase1\_B\_baseline\_seed2} & Spaceship & Baseline    &  745 & 155 & 161.2 & 2.62 & 1.50 \\
    \texttt{phase1\_B\_baseline\_seed3} & Spaceship & Baseline    &  601 & 120 & 132.4 & 2.38 & 1.12 \\
    \texttt{phase1\_B\_baseline\_seed4} & Spaceship & Baseline    &  536 & 101 & 127.1 & 2.57 & 1.00 \\
    \texttt{phase2\_B\_within\_seed1}   & Spaceship & Within-comp &  781 & 116 & 135.8 & 2.12 & 1.12 \\
    \texttt{phase2\_B\_within\_seed2}   & Spaceship & Within-comp &  756 & 120 & 112.1 & 2.00 & 1.25 \\
    \texttt{phase2\_B\_within\_seed3}   & Spaceship & Within-comp & 1253 & 130 & 140.3 & 2.38 & 1.12 \\
    \texttt{phase2\_B\_within\_seed4}   & Spaceship & Within-comp &  925 & 162 & 162.1 & 2.50 & 1.75 \\
    \bottomrule
  \end{tabular}
\end{table*}

\begin{table*}[t]
  \centering
  \small
  \caption{\textbf{Appendix~C: Pairwise Statistical Tests.} Baseline vs.\ Within-comp. Mann-Whitney~U test (two-sided), $n=4$ per group unless noted. 95\%~CI bootstrapped (10\,000 resamples) on the mean difference (Baseline $-$ Within-comp). Cohen's $d = (\mu_{\text{base}} - \mu_{\text{within}}) / s_{\text{pooled}}$; negative $d$ means within-comp is higher. \textbf{Bold}: $p < 0.05$ (only result reaching conventional significance). $^\dagger$~NOMAD SOTA gain: Baseline $n=3$ (seed~2 best in loop~0).}
  \label{tab:a3_stat_tests}
  \begin{tabular}{p{3.6cm}p{2.2cm}p{2.2cm}rrp{2.6cm}p{1.4cm}}
    \toprule
    \textbf{Metric} & \textbf{Baseline mean (SD)} & \textbf{Within-comp mean (SD)} & \textbf{Cohen's $d$} & \textbf{$p$ (MW)} & \textbf{95\% CI} & \textbf{Winner} \\
    \midrule
    \multicolumn{7}{l}{\textit{\textbf{NOMAD 2018}}}\\
    \midrule
    Best metric                      & 0.0608 (0.0038) & 0.0632 (0.0027) & $-$0.75 & 0.200 & [$-$0.0063,\,$+$0.0014] & Baseline \\
    First-loop metric                & 0.0641 (0.0063) & 0.0653 (0.0030) & $-$0.23 & 0.886 & [$-$0.0071,\,$+$0.0048] & Baseline \\
    SOTA gain (abs)$^\dagger$        & 0.0044 (0.0025) & 0.0009 (0.0009) & $+$1.99 & 0.114 & [$+$0.0009,\,$+$0.0056] & Baseline \\
    SOTA gain (rel)$^\dagger$        & 6.50\% (3.60\%) & 1.41\% (1.37\%) & $+$2.03 & 0.114 & [$+$1.51\%,\,$+$8.30\%]  & Baseline \\
    Loops to 1st improvement         & 2.00 (2.00)     & 2.25 (1.89)     & $-$0.13 & 0.739 & [$-$2.50,\,$+$2.00]      & Baseline \\
    Total run.\ evo-loops            & 19.0 (2.8)      & 20.0 (0.8)      & $-$0.48 & 1.000 & [$-$3.8,\,$+$1.2]        & Baseline \\
    Total cod.\ evo-loops            & 10.2 (2.2)      &  9.5 (0.6)      & $+$0.46 & 0.882 & [$-$1.0,\,$+$2.8]        & Within-comp \\
    Retention rate                   & 0.429 (0.285)   & 0.438 (0.161)   & $-$0.04 & 1.000 & [$-$0.281,\,$+$0.263]    & Within-comp \\
    Prompt tokens (k)                & 877 (226)       & 913 (30)        & $-$0.22 & 0.486 & [$-$234,\,$+$170]         & Baseline \\
    Completion tokens (k)            & 150 (21)        & 144 (6)         & $+$0.40 & 0.686 & [$-$13,\,$+$24]           & Within-comp \\
    Duration (min)                   & 159.8 (29.7)    & 159.5 (16.2)    & $+$0.01 & 0.886 & [$-$30.7,\,$+$26.4]       & Within-comp \\
    \midrule
    \multicolumn{7}{l}{\textit{\textbf{Spaceship Titanic}}}\\
    \midrule
    Best metric                      & 0.8098 (0.0042) & 0.8147 (0.0024) & $-$1.41 & 0.114 & [$-$0.0091,\,$-$0.0009]  & Within-comp \\
    First-loop metric                & 0.8004 (0.0169) & 0.8031 (0.0019) & $-$0.23 & 0.486 & [$-$0.0196,\,$+$0.0078]  & Within-comp \\
    SOTA gain (abs)                  & 0.0095 (0.0131) & 0.0116 (0.0013) & $-$0.23 & 0.343 & [$-$0.0098,\,$+$0.0109]  & Within-comp \\
    SOTA gain (rel)                  & 1.21\% (1.69\%) & 1.45\% (0.16\%) & $-$0.19 & 0.343 & [$-$1.23\%,\,$+$1.46\%]  & Within-comp \\
    Loops to 1st improvement         & 1.25 (0.50)     & 1.25 (0.50)     &    0.00 & 1.000 & [$-$0.50,\,$+$0.50]       & --- \\
    Total run.\ evo-loops            & 18.5 (2.1)      & 18.0 (1.8)      & $+$0.26 & 0.884 & [$-$1.8,\,$+$2.8]         & Within-comp \\
    Total cod.\ evo-loops            & 10.2 (2.8)      & 10.5 (2.4)      & $-$0.10 & 0.882 & [$-$3.5,\,$+$2.5]         & Baseline \\
    Retention rate                   & 0.415 (0.229)   & 0.562 (0.161)   & $-$0.74 & 0.306 & [$-$0.366,\,$+$0.103]     & Within-comp \\
    \textbf{Prompt tokens (k)}       & \textbf{625 (87)} & \textbf{929 (229)} & $\mathbf{-1.76}$ & $\mathbf{0.029}$ & [$-$527,\,$-$121] & \textbf{Baseline} \\
    Completion tokens (k)            & 127 (23)        & 132 (21)        & $-$0.24 & 1.000 & [$-$33,\,$+$20]            & Baseline \\
    Duration (min)                   & 140.6 (15.0)    & 137.6 (20.5)    & $+$0.17 & 1.000 & [$-$18.0,\,$+$24.9]        & Within-comp \\
    \bottomrule
  \end{tabular}
\end{table*}

\begin{table*}[t]
  \centering
  \small
  \caption{\textbf{Appendix~D: Seed Knowledge Base Composition.} Composition of the seed knowledge base (\texttt{seed\_union\_AB}) by model type. The KB contains 30 cases from NOMAD and 30 from Spaceship Titanic (model-type breakdown not available at competition level for the fused KB). All 60 cases are labelled {\normalfont\textit{outcome = success}}; 2 failure patterns recorded.}
  \label{tab:a4_seed_kb}
  \begin{tabular}{lr}
    \toprule
    \textbf{Model type} & \textbf{Cases} \\
    \midrule
    xgboost & 24 \\
    lightgbm & 13 \\
    ensemble (random\_forest + xgboost) & 9 \\
    ensemble (lightgbm + xgboost) & 5 \\
    random\_forest & 4 \\
    ensemble (gradient\_boosting + xgboost) & 2 \\
    catboost & 2 \\
    ensemble (catboost + lightgbm + xgboost) & 1 \\
    \midrule
    \textbf{Total} & \textbf{60} \\
    \bottomrule
  \end{tabular}
\end{table*}

\begin{table*}[t]
  \centering
  \small
  \caption{\textbf{Appendix~E: Per-Run Knowledge Base Statistics.} \textbf{Cases} = total cases in KB at end of run. \textbf{New} = cases added during the run (within-comp: Cases $-$ 60 seed cases; baseline: all cases). \textbf{Fail.}\ = failure patterns recorded. \textbf{Reuse} = total retrieval events from KB. \textbf{Succ.}\ = retrieval events classified as successful. \textbf{Uniq.}\ = distinct case IDs retrieved at least once. Baseline runs have Reuse $= 0$ by design (retrieval disabled).}
  \label{tab:a5_kb_stats}
  \begin{tabular}{lllrrrrrr}
    \toprule
    \textbf{Run ID} & \textbf{Comp.} & \textbf{Cond.} & \textbf{Cases} & \textbf{New} & \textbf{Fail.} & \textbf{Reuse} & \textbf{Succ.} & \textbf{Uniq.} \\
    \midrule
    \texttt{phase1\_A\_baseline\_seed1} & NOMAD & Baseline    &  8 &  8 & 0 &  0 &  0 & 0 \\
    \texttt{phase1\_A\_baseline\_seed2} & NOMAD & Baseline    &  7 &  7 & 1 &  0 &  0 & 0 \\
    \texttt{phase1\_A\_baseline\_seed3} & NOMAD & Baseline    &  7 &  7 & 0 &  0 &  0 & 0 \\
    \texttt{phase1\_A\_baseline\_seed4} & NOMAD & Baseline    &  8 &  8 & 0 &  0 &  0 & 0 \\
    \texttt{phase2\_A\_within\_seed1}   & NOMAD & Within-comp & 68 &  8 & 2 & 31 & 15 & 6 \\
    \texttt{phase2\_A\_within\_seed2}   & NOMAD & Within-comp & 68 &  8 & 2 & 27 & 11 & 6 \\
    \texttt{phase2\_A\_within\_seed3}   & NOMAD & Within-comp & 68 &  8 & 2 & 30 & 14 & 9 \\
    \texttt{phase2\_A\_within\_seed4}   & NOMAD & Within-comp & 67 &  7 & 3 & 29 & 13 & 8 \\
    \midrule
    \texttt{phase1\_B\_baseline\_seed1} & Spaceship & Baseline    &  8 &  8 & 0 &  0 &  0 & 0 \\
    \texttt{phase1\_B\_baseline\_seed2} & Spaceship & Baseline    &  8 &  8 & 0 &  0 &  0 & 0 \\
    \texttt{phase1\_B\_baseline\_seed3} & Spaceship & Baseline    &  7 &  7 & 1 &  0 &  0 & 0 \\
    \texttt{phase1\_B\_baseline\_seed4} & Spaceship & Baseline    &  7 &  7 & 0 &  0 &  0 & 0 \\
    \texttt{phase2\_B\_within\_seed1}   & Spaceship & Within-comp & 68 &  8 & 2 & 30 & 14 & 8 \\
    \texttt{phase2\_B\_within\_seed2}   & Spaceship & Within-comp & 68 &  8 & 2 & 31 & 15 & 9 \\
    \texttt{phase2\_B\_within\_seed3}   & Spaceship & Within-comp & 68 &  8 & 2 & 29 & 13 & 10 \\
    \texttt{phase2\_B\_within\_seed4}   & Spaceship & Within-comp & 68 &  8 & 2 & 29 & 13 & 8 \\
    \bottomrule
  \end{tabular}
\end{table*}

\begin{table*}[t]
  \centering
  \small
  \caption{\textbf{Appendix~F: Per-Run CBR Heuristic Reuse Statistics} (within-comp runs only). \textbf{Events} = logged retrieval events. \textbf{Uniq.} = distinct case IDs retrieved at least once. \textbf{emb} = embedding cosine similarity (threshold 0.82). \textbf{fp} = code fingerprint similarity (threshold 0.10). All 108 events met both similarity thresholds (heuristic reuse check); no negative outcomes recorded.}
  \label{tab:a6_causal_reuse}
  \begin{tabular}{llrrrrrr}
    \toprule
    \textbf{Run ID} & \textbf{Comp.} & \textbf{Events} & \textbf{Uniq.} & \textbf{emb mean (SD)} & \textbf{fp mean (SD)} & \textbf{fp min} & \textbf{fp max} \\
    \midrule
    \texttt{phase2\_A\_within\_seed1} & NOMAD     & 15 &  6 & 0.894 (0.011) & 0.275 (0.217) & 0.100 & 0.750 \\
    \texttt{phase2\_A\_within\_seed2} & NOMAD     & 11 &  4 & 0.889 (0.037) & 0.198 (0.110) & 0.102 & 0.499 \\
    \texttt{phase2\_A\_within\_seed3} & NOMAD     & 14 &  8 & 0.909 (0.034) & 0.342 (0.225) & 0.117 & 0.761 \\
    \texttt{phase2\_A\_within\_seed4} & NOMAD     & 13 &  5 & 0.887 (0.040) & 0.261 (0.155) & 0.101 & 0.592 \\
    \midrule
    \texttt{phase2\_B\_within\_seed1} & Spaceship & 14 &  7 & 0.861 (0.019) & 0.275 (0.115) & 0.113 & 0.462 \\
    \texttt{phase2\_B\_within\_seed2} & Spaceship & 15 &  8 & 0.868 (0.021) & 0.364 (0.160) & 0.163 & 0.711 \\
    \texttt{phase2\_B\_within\_seed3} & Spaceship & 13 &  7 & 0.881 (0.033) & 0.319 (0.191) & 0.101 & 0.757 \\
    \texttt{phase2\_B\_within\_seed4} & Spaceship & 13 &  5 & 0.872 (0.024) & 0.380 (0.158) & 0.191 & 0.762 \\
    \midrule
    \textbf{All within-comp} & --- & \textbf{108} & \textbf{34} & \textbf{0.882 (0.031)} & \textbf{0.305 (0.176)} & \textbf{0.100} & \textbf{0.762} \\
    \bottomrule
  \end{tabular}
\end{table*}

\begin{figure*}[t]
  \centering
  \usetikzlibrary{arrows.meta, positioning, fit, backgrounds, calc}

  \begin{tikzpicture}[
    font=\small,
    % ── Phase box styles ─────────────────────────────────────
    phase/.style={
      draw, rounded corners=4pt,
      minimum width=92mm, minimum height=11mm,
      align=left, text width=87mm,
      inner sep=4pt, line width=0.5pt,
    },
    cbr/.style  ={phase, fill=black!9,  draw=black!65},
    orig/.style ={phase, fill=black!3,  draw=black!35},
    % ── Component box styles ─────────────────────────────────
    comp/.style={
      draw, rounded corners=4pt,
      minimum width=50mm, align=left,
      text width=46mm, inner sep=4pt,
      line width=0.5pt,
    },
  kb/.style={comp, fill=white, draw=black!70, line width=0.9pt},
  ft/.style={comp, fill=white, draw=black!70, line width=0.9pt},
  qg/.style={comp, fill=white, draw=black!70, line width=0.9pt},
    % ── Arrow styles ─────────────────────────────────────────
    arr/.style ={-{Stealth[length=4pt,width=3.5pt]},
                line width=0.65pt, black!80},
    arrd/.style={-{Stealth[length=3.5pt,width=3pt]},
                line width=0.55pt, dashed, black!60},
    albl/.style={font=\tiny\itshape, inner sep=1.5pt,
                fill=white, fill opacity=0.92, text opacity=1},
  ]

  % ── Phase boxes (left column) ────────────────────────────
  \node[cbr] (p1) {%
    \textbf{Phase 1 \textendash{} Hypothesis Generation}%
    \hfill{\scriptsize\textit{CBR override}}\\[2pt]
    \scriptsize
    Retrieve relevant cases and failure patterns from KB\,/\,FT;
    enrich generated hypothesis with CBR context%
  };
  \node[cbr, below=4mm of p1] (p2) {%
    \textbf{Phase 2 \textendash{} Code Generation}%
    \hfill{\scriptsize\textit{CBR override}}\\[2pt]
    \scriptsize
    Inject retrieved code snapshots as reference context;
    run CoSTEER debug-iterate sub-loop \textit{(unchanged)}%
  };
  \node[orig, below=4mm of p2] (p3) {%
    \textbf{Phase 3 \textendash{} Execution}%
    \hfill{\scriptsize\textit{unchanged}}\\[2pt]
    \scriptsize
    Execute generated code in isolated environment; collect outputs%
  };
  \node[orig, below=4mm of p3] (p4) {%
    \textbf{Phase 4 \textendash{} Feedback}%
    \hfill{\scriptsize\textit{unchanged}}\\[2pt]
    \scriptsize
    LLM quality judge evaluates result; produces structured feedback%
  };
  \node[cbr, below=4mm of p4] (p5) {%
    \textbf{Phase 5 \textendash{} CBR Retention}%
    \hfill{\scriptsize\textit{CBR override}}\\[2pt]
    \scriptsize
    Apply five-gate quality filter; retain successful cases in KB,
    log failures to FT, update heuristic reuse counters%
  };

  % ── Sequential phase arrows ──────────────────────────────
  \draw[arr] (p1.south) -- (p2.north);
  \draw[arr] (p2.south) -- (p3.north);
  \draw[arr] (p3.south) -- (p4.north);
  \draw[arr] (p4.south) -- (p5.north);

  % ── Left loop-back rail ──────────────────────────────────
  \draw[arr] (p5.west) --++(-8mm,0) coordinate(lrail)
    |- node[font=\tiny, text=black!40,
            rotate=90, anchor=south]
        {\textit{next loop}} (p1.west);

  % ── Component boxes (right column, 12mm gap from phases) ─
  \node[kb, right=12mm of p1.north east, anchor=north west] (kb) {%
    \textbf{Knowledge Base (KB)}\\[2pt]
    \raggedright\hyphenpenalty=10000\scriptsize
    Persistent store of quality-filtered cases\\[1pt]
    \textbf{Each case:} problem signature, solution
    plan, code snapshot, quality metrics\\[1pt]
    \textbf{Ops:} retrieve\,/\,add\,/\,heuristic reuse check%
  };
  \node[ft, below=5mm of kb] (ft) {%
    \textbf{Failure Tracker (FT)}\\[2pt]
    \raggedright\hyphenpenalty=10000\scriptsize
    Embedding index of failed runs;\\
    warns about known bad methods%
  };
  \node[qg, below=5mm of ft] (qg) {%
    \textbf{Quality Gate (QG)}\\[2pt]
    \raggedright\scriptsize
    \textbf{G1}~Execution success\\
    \textbf{G2}~Metric extractable\\
    \textbf{G3}~Improvement over SOTA\\
    \textbf{G4}~Novelty (deduplication)\\
    \textbf{G5}~LLM enrichment%
  };

  % ── Cross-column arrows ──────────────────────────────────

  % (a) KB -> Phase 1: retrieve — erst vertikal auf P1-Höhe, dann horizontal rein
  \coordinate (p1eTop) at ($(p1.east)+(0, 3mm)$);
  \coordinate (kbWTop) at ($(kb.west) +(0, 3mm)$);
  \draw[arr]
    (kbWTop)
    -- (kbWTop |- p1eTop)
    -- node[albl, above]{retrieve} (p1eTop);

  % (b) KB -> Phase 2: retrieve — gerade runter, dann horizontal rein
  \coordinate (p2eA) at ($(p2.east)+(0, 2mm)$);
  \draw[arrd]
    (kb.south)
    -- (kb.south |- p2eA)
    -- node[albl, above]{retrieve} (p2eA);

  % (c) FT -> Phase 1: failures (dashed) — unverändert, kommt bereits horizontal an
  \coordinate (gapX)    at ($(p1.east)+(5mm, 0)$);
  \coordinate (p1eLow)  at ($(p1.east)+(0,-3mm)$);
  \coordinate (ftToGap) at (ft.west -| gapX);
  \coordinate (gapToP1) at (gapX |- p1eLow);
  \coordinate (failMid) at ($(ftToGap)!0.5!(gapToP1)$);
  \draw[arrd] (ft.west) -- (ftToGap) -- (gapToP1) -- (p1eLow);
  \node[albl, left] at (failMid) {failures};

  % (d) QG -> Phase 5: evaluate
  \coordinate (p5eEval) at ($(p5.east)+(0,-2mm)$);   % war: +2mm
  \coordinate (qgW)     at ($(qg.west) +(0,-2mm)$);  % war: +2mm
  \draw[arr]
    (qgW)
    -- (qgW |- p5eEval)
    -- node[albl, above]{evaluate} (p5eEval);

  % (e) Phase 5 -> KB: add case (solid, right rail)
  \coordinate (railR)  at ($(kb.east)+(8mm,0)$);
  \coordinate (p5eTop) at ($(p5.east)+(0, 2.5mm)$);
  \coordinate (kbArr)  at ($(kb.east)+(0,-2mm)$);
  \draw[arr]
    (p5eTop) -- (p5eTop -| railR)
    -- node[albl, left, pos=0.62]{add case}
      (railR |- kbArr) -- (kbArr);

  % (f) Phase 5 -> FT: log failure (dashed, eigene Schiene weiter innen)
  \coordinate (railR2) at ($(kb.east)+(4mm,0)$);   % <-- Wert anpassen: kleiner = weiter innen
  \coordinate (p5eLow) at ($(p5.east)+(0,-2.5mm)$);
  \coordinate (ftArr)  at ($(ft.east)+(0, 2mm)$);
  \draw[arrd]
    (p5eLow) -- (p5eLow -| railR2)
    -- node[albl, left, pos=0.35]{log failure}
      (railR2 |- ftArr) -- (ftArr);

  % ── Dashed loop outline (wraps phase boxes only) ─────────
  \begin{scope}[on background layer]
    \node[draw=black!22, dashed, rounded corners=8pt,
          inner xsep=9mm, inner ysep=6mm,
          fit=(p1)(p2)(p3)(p4)(p5)] (loopbox) {};
  \end{scope}
  \node[font=\scriptsize, text=black!40,
        above=2mm of loopbox.north, anchor=south]
    {\textit{CBR-augmented R\&D-Agent loop (one iteration)}};

  % ── Legend (single centered row across full figure width) ─
  % figMidBot = midpoint between lrail and railR at 8mm below loopbox.south
  \coordinate (figMidBot) at
    ($(lrail |- loopbox.south)!0.5!(railR |- loopbox.south)+(0,-8mm)$);
  \node[anchor=north] at (figMidBot) {%
    \setlength{\tabcolsep}{4pt}%
    \begin{tabular}{@{}c@{\,}l@{\quad}c@{\,}l@{\quad}c@{\,}l@{\quad}c@{\,}l@{}}
      \tikz[baseline=3pt]
        \node[cbr, minimum width=9mm, minimum height=4.5mm,
              inner sep=0, text width=0]{};
      & \small CBR override
      & \tikz[baseline=3pt]
        \node[orig, minimum width=9mm, minimum height=4.5mm,
              inner sep=0, text width=0]{};
      & \small Unchanged phase
      & \tikz[baseline=3.5pt]{\draw[arr,black!70](0,0)--(13mm,0);}
      & \small Data\,/\,control flow
      & \tikz[baseline=3.5pt]{\draw[arrd,black!60](0,0)--(13mm,0);}
      & \small Optional\,/\,indirect
    \end{tabular}%
  };

  \end{tikzpicture}
  \caption{\textbf{Appendix~G: CBR-augmented R\&D-Agent loop (one iteration).}
  Dark grey boxes mark the three phases overridden by the CBR layer
  (Phases~1, 2, 5); light grey boxes mark unchanged R\&D-Agent phases
  (Phases~3, 4).
  Phase~1 performs two-stage CBR retrieval and injects retrieved cases and
  Failure-Tracker patterns into the hypothesis; Phase~2 appends code snapshots
  to the coding prompt; Phase~5 applies the five-gate Quality Gate and updates
  heuristic reuse counters.
  See Algorithm~\ref{alg:cbr_loop} for the corresponding pseudocode.
  \textbf{KB} = persistent Knowledge Base (case metadata, code snapshots,
  vector index); \textbf{FT} = Failure Tracker; \textbf{QG} = Quality Gate
  (gates G1--G5). Solid arrows = data\,/\,control flow; dashed arrows =
  optional\,/\,indirect paths.}
  \label{fig:rdagent_loop}
\end{figure*}
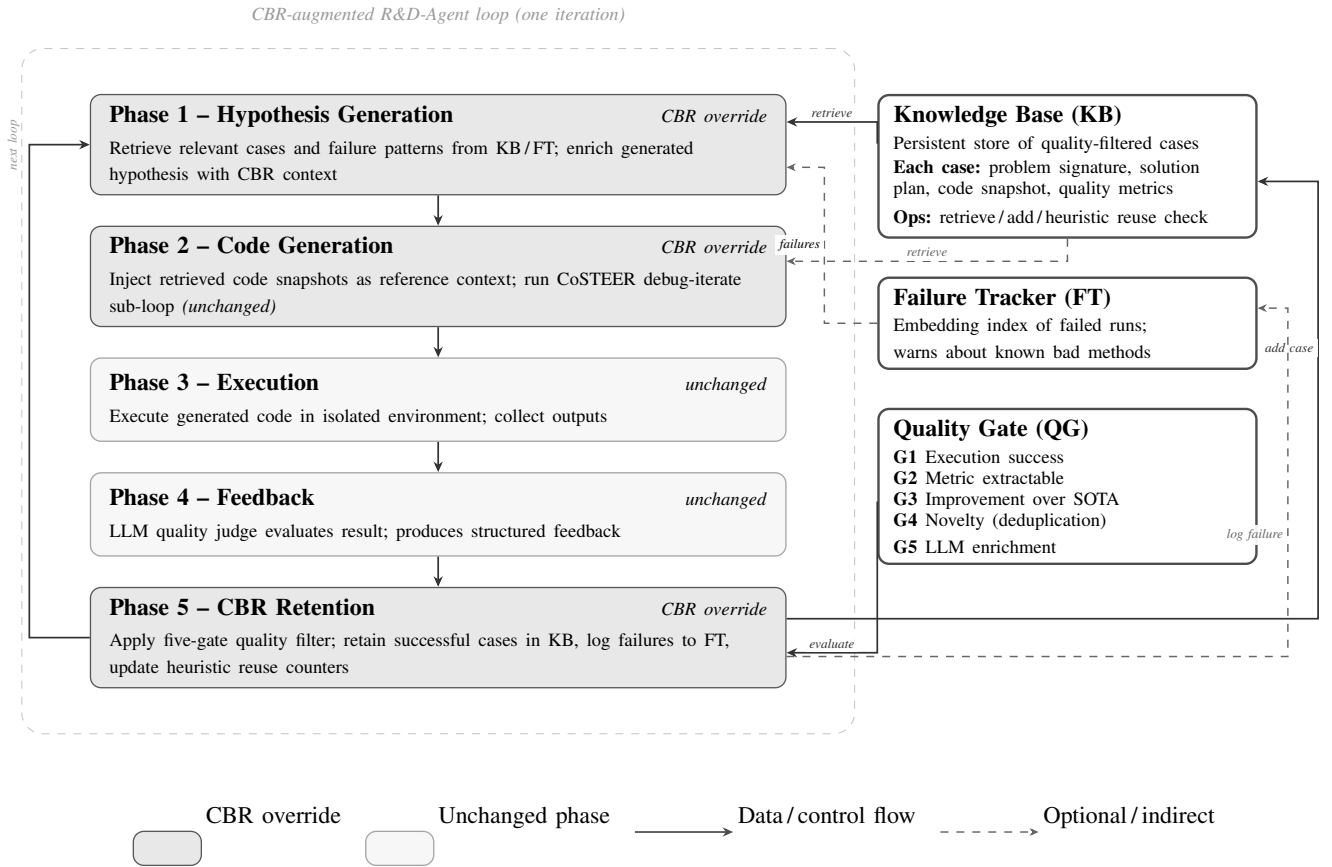

\end{document}